\documentclass[conference]{IEEEtran}
\usepackage{graphicx}
\usepackage{cite}
\usepackage{comment}
\usepackage{amsmath,amssymb,amsfonts}
\usepackage{xcolor}
\usepackage{wrapfig,paralist}
\usepackage{lscape}
\usepackage{rotating}
\usepackage{epstopdf}
\usepackage{color,balance}
\usepackage{amsthm}
\usepackage{savesym}
\savesymbol{state}
\usepackage{tengwarscript}
\usepackage{grumble}
\usepackage[hyphens]{url}
\usepackage{algorithmic}
\usepackage[linesnumbered,ruled,noend]{algorithm2e}
\usepackage{array}

\usepackage{hyperref}
\usepackage[hyphenbreaks]{breakurl}
\usepackage[noabbrev, capitalise]{cleveref}

\newtheorem{theorem}{Theorem}[section]

\newtheorem{definition}[theorem]{Definition}

\newcommand{\nobs}{\mathcal{O}}

\newcommand{\obs}[1]{\mathcal{O}(#1)}
\newcommand{\state}[1]{\mathcal{R}(#1)}
\newcommand{\del}{t_{del}}
\newcommand{\create}{t_0}

\newcommand{\upDur}{\Delta t_u}
\newcommand{\downDur}{\Delta t_d}

\newcommand{\ccdfUp}[1]{\overline{F_{T_u}}(#1) }
\newcommand{\ccdfDo}[1]{\overline{F_{T_d}}(#1) }

\newcommand{\fUp}[1]{f_{T_u}(#1)}

\newcommand{\system}{\textsl{Lethe}}

\newcommand{\decisionThres}{\theta}
\newcommand{\decisionThresEst}{{\theta^*}}

\renewcommand{\paragraph}[1]{\vspace{0.3mm}\noindent \textbf{#1.\ }}

\newcommand{\post}{post}
\newcommand{\posts}{posts}

\newcommand{\platform}{platform}

\begin{document}

	\title{Forgetting the Forgotten with \system\\{\LARGE Concealing Content Deletion from Persistent Observers}}
	
		\author{
			\IEEEauthorblockN{Mohsen Minaei}
			\IEEEauthorblockA{Purdue University\\
			mohsen@purdue.edu
		}
			\and
			\IEEEauthorblockN{Mainack Mondal}
			\IEEEauthorblockA{University of Chicago\\
			mainack@uchicago.edu
		}
			\and
			\IEEEauthorblockN{Patrick Loiseau}
			\IEEEauthorblockA{ Univ. Grenoble Alpes, CNRS,\\Inria, Grenoble INP, LIG\\ 
			patrick.loiseau@inria.fr
		}
			\and
			\IEEEauthorblockN{Krishna Gummadi}
			\IEEEauthorblockA{MPI-SWS \\
			gummadi@mpi-sws.org
			}
			\and
			\IEEEauthorblockN{Aniket Kate}
			\IEEEauthorblockA{Purdue University\\
			aniket@purdue.edu
			}

		}

	\IEEEoverridecommandlockouts

	\maketitle

	\begin{abstract}
Most social platforms offer mechanisms allowing users to delete their posts, and a significant fraction of users exercise this right to be forgotten.
However, ironically, users' attempt to reduce attention to sensitive posts via deletion, in practice, attracts unwanted attention from stalkers specifically to those (deleted) posts. 
Thus, deletions may leave users more vulnerable to attacks on their privacy in general. 
Users hoping to make their posts forgotten face a ``damned if I do, damned if I don't'' dilemma. 
Many are shifting towards ephemeral social platform like Snapchat, which will deprive us of important user-data archival.
In the form of intermittent withdrawals, 
we present, \system, a novel solution to this problem of (really) forgetting the forgotten.
If the next-generation social platforms are willing to give up the uninterrupted availability 
of {\em non-deleted} posts by a very small fraction,  
\system\ provides privacy to the deleted posts over long durations. 
In presence of \system{}, an adversarial observer becomes unsure if some posts are permanently deleted or just temporarily withdrawn by \system{}; at the same time, the adversarial observer is overwhelmed by a large number of falsely flagged undeleted posts.
To demonstrate the feasibility and performance of \system{}, 
we analyze large-scale real data about users' deletion over Twitter  
and thoroughly investigate how to choose time duration distributions for alternating between temporary withdrawals and resurrections of non-deleted posts. 
We find a favorable trade-off between privacy, availability 
and adversarial overhead in different settings for users exercising their right to delete.
We show that, even against an ultimate adversary with an uninterrupted access to the entire platform,    
\system{} offers deletion privacy for up to $3$ months from the time of deletion,
while maintaining content availability as high as $95\%$ and keeping the adversarial precision to $20\%$.

\end{abstract}

	\maketitle

\section{Introduction}
\noindent People freely open up about their personal life and opinions on online social platforms (e.g., Facebook, Twitter) today. The shared information remains available on these platforms (to intended recipients as well as unintended observers) and is archived by archival services until (and if) the information is eventually deleted (or confined) by its creator. This long-term exposure of the shared data raises numerous
longitudinal privacy concerns~\cite{bauer-2013-postanachronism,ayalon-2013-retrospective,mondal-2016-longitudinal-exposure} for the users: not only celebrities but non-celebrities get regularly harassed and blackmailed by data scavengers, who stalk their victims to identify sensitive content from the shared data. Nevertheless, sensitivity of a post is relative; it varies from person to person, and also with life events and time in general. Thus, effective (high precision and recall) mining of available large-scale data to find suitable victims is not always
feasible for the scavengers.

The task should have become more difficult as platforms and Internet archives honor users' request to delete their data. However, these deletions actually leave the users more vulnerable to the scavengers who can now focus only on the 
withdrawn posts to find sensitive contents.\footnote{Closely associated phenomenon, ``Streisand effect,'' suggests that an attempt to hide some information has the unintended consequence of bringing
	particular attention of public to it.} Indeed, we found this problem associated with content deletions to be very practical---today multiple web services find and hoard deleted content across different social platforms. 
Politwoops~\cite{Politwoops} 
for Twitter, ReSavr~\cite{resavr} and Uneddit~\cite{uneddit} for Reddit, StackPrinter-Deleted~\cite{StackPrinter} for Stack overflow,
and YouTomb~\cite{YouTomb} for Youtube are some of the prominent examples.
In fact, Politwoops archived more than 1.1 million deleted tweets by 10,404 politicians, 
around the world in 2015~\cite{politwoops_archive}, and by August 2017 Uneddit serves more than 942 million deleted Reddit comments. 
These services can 
enable attackers to specifically mine deleted posts of users for nefarious purposes.

This large-scale identification and hoarding of deleted content from social sites and archives pose a serious violation of ``Right to be Forgotten'' and the ill-effects of this phenomena on our social behavior will be far reaching. For example, in one case, singer Ed Sheeran's deletion of a tweet from 2011 was found and widely publicized in media~\cite{ed_shereen_buzzfeed} leading to his brief disappearance from Twitter. In another case, an SNL cast member's deletion of racist tweets back in September 2016~\cite{SNL} were tracked by third parties and subsequently publicized. Not only celebrities but normal users also fell prey to this phenomenon when links delisted by Google in Europe (to honor Right to be Forgotten requests) were identified, publicized and scrutinized by media~\cite{xue2016right}. In general, the users today are extremely vulnerable due to the fact that, whatever content they delete (ironically, to protect their privacy) will possibly be identified, dissected and abused.

In spite of this threat, not surprisingly, without any better alternatives available, information exposure control in the form of deletions  still remains a common phenomenon on the social platforms; Mondal et al.~\cite{mondal-2016-longitudinal-exposure} observe that a significant fraction ($\sim$35\%) of all Twitter users  have now deleted or confined (i.e., made private) their public Twitter posts made in 2009. Consequently, as any persistent onlooker can keep track of such changes and go after the deleted posts, 
users aiming to make observers forget their posts are left with a ``damned if I do, damned if I don't'' dilemma. This paper aims to provide a solution to the problem.

A trivial solution is to make users {\em not} publish sensitive content in the first place; but this 
is infeasible even for extremely careful users as the sensitivity of shared data  changes drastically and unpredictably with time and life events. 
A growing number of users have now shifted to ephemeral social platforms such as Snapchat~\cite{snapchat}, 
where everything gets deleted 
in a premeditated fashion.
However, given the huge historical, cultural, and economical value of user-generated data,
it is extremely unlikely that most next-generation social or archival platforms will adapt to this model.

This leaves us with a hard research question: {\em can we offer an alternative to the next-generation social or archival platforms that achieves the best properties of both deleting everything (i.e., privacy)  and keeping an archive of posts and events (i.e., availability)?} The aim of this work is to answer this question affirmatively
and develop a privacy mechanism that retains the archival values of posted content and still allows deletions while providing deniability and protection to the users after some time of deletion, i.e., those deletions will not be immediately discernible to even persistent onlookers.

\paragraph{A simple-yet-drastic proposal} We offer a simple-yet-drastic proposal towards mitigating the problem of concealing content deletions in presence of persistent observers while maintaining 
high availability of archived content. In our proposed system, \system\footnote{In Greek mythology, 
	\system\ was the river of forgetfulness: all those who drank from it experienced complete forgetfulness. 
	The word \system\ also means oblivion, forgetfulness, or concealment.}, 
we very conservatively assume that the adversary has complete access to the archival platform and can view any post. We presume the platform administrator is working with the data creator (or owner) to protect the privacy of deletions. \system{} employs an {\em intermittent withdrawal mechanism} that protects privacy using two public, infinite-support time distributions---one we call the up (or online) distribution and the second is called down (or offline) distribution. 
Just before publishing a post, \system{} samples a time duration from the up distribution and for that time duration makes that post available (i.e., visible) to everyone. After the up duration passes \system{} takes an instance from the down distribution and for that time duration hides the post from viewers.

In the same way, \system{} continues to toggle between the up and down durations as long as the post has not been deleted or its privacy preference has not changed. 
Since \system{} also hides non-deleted posts, 
it will be confusing for the adversary to distinguish whether a post is hidden by \system{} or deleted by the owner.

\paragraph{Contributions}
We make four key contributions.

Firstly, to the best of our knowledge, this is the first systematic study of the problem with content deletion in the presence of persistent onlookers. 
We formalize the problem with content deletion in the presence of a very powerful adversary who can take snapshots of the whole platform at any point in time.
We define and analytically quantify the necessary security notions: \textit{privacy}--likelihood ratio of a post deleted or not at any particular time, \textit{availability}--fraction of time the posts are visible  and \textit{adversarial overhead}--adversary's precision on detecting deleted posts.
Based on our formalization, we propose and evaluate a novel scheme, \system{}, to provide privacy for users' deletions.

Secondly, we show that privacy is correlated with the up and down distributions: 
(i) inversely proportional to the hazard rate of up distribution, 
and (ii) inversely proportional to the complementary cumulative distribution function (CCDF) of down distribution. 
Moreover, we show that by picking geometric and negative binomial distributions as the up and down distribution,
not only we achieve good privacy guarantees, but our notion of privacy is simplified to a \textit{decision threshold} period---duration an adversary is willing to wait before identifying a (hidden) post in a down period as deleted.

Thirdly, we present the trade-offs between the notions mentioned above using data from Twitter. 
We show that in the case of 95\% content availability, 
the adversary, with an uninterrupted access to the entire platform, 
will have a precision value associated with adversarial overhead below 20\%  
even when a post has been down for more than 90 days. 
In the case of a more forbearing adversary that has a decision threshold of 180 days, the precision will only increase to 35\%.
However, the system administrator can reduce the availability of the system by a small fraction and set it to 90\%, which drops the adversary's precision back to 20\%.
For a large-scale system such as Twitter, with trillions of tweets, even precision of 80\% can result in a significant overhead for the adversary (investigating 20 million  non-deleted tweets falsely marked as deleted each day). 

Finally, we evaluate the effect of our scheme on Twitter dataset to
show the feasibility of \system{} in a real-world scenario.
We show that our proposal, 
while maintaining a trade-off between availability and privacy, 
also allows interactions in the system without much interruption. 
Specifically, leveraging real-world interaction data from Twitter we show that, by applying \system{} 
the utility (i.e. user interactions with posts) remains above 99\%  even when content availability is 85\%. 

\paragraph{Applicability of \system}
Users of platforms such as Twitter or Facebook are accustomed to uninterrupted availability of their uploaded/archived data. 
Any loss of availability (even if loss is small) may be unacceptable to some users, and
such platforms can introduce \system{} as an optional feature (providing an opt-out option for the users as well as applying \system{} only to the posts that are at least days or weeks old) if they find that some of their users demand privacy for their deletions. 
Nevertheless, we primarily envision \system{} for the next-generation social or archival platforms, 
where, unlike current ephemeral platforms like Snapchat, the users expect to have an archive of old memories without the fear of others breaking privacy for their deletions.

Moreover, high availability system like \system{} will be less
effective against an adversary that devotes time and resources on 
a particular user such as the case of unearthing Ed Sheeran's deletions~\cite{ed_shereen_buzzfeed}.
Nevertheless, compared to the state-of-the-art, \system\ raises the bar significantly:
it not only offers deniability to the celebrity for at least a few arguably important weeks, 
but also significantly increases the stalker's workload.

	\section{Context and Motivation}\label{sec:background}

\noindent In this section, we explain how the current systems handle content deletions and what are the problems with those privacy methodologies. Furthermore, 
we motivate our proposal towards achieving privacy and availability simultaneously.

\paragraph{User-initiated spontaneous deletions}
One of the widely employed form of content deletion today is user-initiated deletion; 
i.e., system operators remove content when the owners explicitly asked them to do so. 
Almost all real world social data sharing \platform{}s today (e.g., Facebook, Twitter or YouTube) 
provide users option to delete their uploaded content. 
Recent studies~\cite{mondal-2016-longitudinal-exposure,
	almuhimedi-2013-deleteTweets} have shown that users extensively use this mechanism to protect the privacy of their 
past content---users delete around 35\% of posts within six years of posting them.  
The European Union (EU) regulation of ``Right to be forgotten''~\cite{xue2016right,weber-2011-rightToForget} which is part of EU General Data Protection Regulation (GDPR)~\cite{GDPR} is also trying to accomplish exactly this same, 
albeit at a much more elaborate scale. They wish to enable users to remove historical data about themselves from multiple systems, 
including removing results from leading search engines. Nevertheless, as we already suggest, 
those deleted content attract unwanted attention~\cite{xue2016right}. 

\paragraph{Premeditated withdrawals} 
Complementary to these user-initiated spontaneous deletions,
a number of {\em premeditated} withdrawal methodologies have been proposed 
and employed today.

Many of those aim to protect content privacy via withdrawing \textit{all} posts
after a predefined viewership or time of posting; we call those the \textit{age-based withdrawals}. 
Recent ephemeral social content sharing sites like Snapchat~\cite{snapchat} or Dust~\cite{dust} are 
prominent examples of age-based withdrawal.
Several academic projects also try to enforce age-based withdrawal in different context; e.g., Vanish~\cite{geambasu-2009-vanish, geambasu-2011-vanishImprove} in distributed hash tables (DHTs), 
EphPub~\cite{ephpub-2011-Castelluccia} and 
\cite{reimann-2012-timedRevocation} using 
DNS caches, and 
Ephemerizer~\cite{perlman-2005-ephemerizer} and its improvement~\cite{nair-2007-newEphemerizer} 
using trusted servers.
A user's inability to a priori predict the right time (or viewership) for her content withdrawal
remains to be the key issue with the age-based withdrawals.
This prevents 
deriving the best possible content availability. 

Mondal et al.~\cite{mondal-2016-longitudinal-exposure} suggest \textit{inactivity-based withdrawal} 
to eliminate the burden on the users to decide expiry times and to facilitate continued discussions around interesting content.  
Unlike in age-based withdrawals, 
where a post is withdrawn after a predefined time or viewership, 
in inactivity-based withdrawal posts can be withdrawn only when it becomes inactive over time, 
i.e., it does not generate any more interactions (e.g., sharing the post by other users). 
Recently proposed Neuralyzer~\cite{Zarras-2016-Neuralyzer} uses a similar concept to maintain the availability of content 
as long as there is sufficient demand for it, and leverages the caching mechanisms of DNS to keep track of the activity. 
A similar idea is also employed 
on sites like 4chan~\cite{hine2017kek, fourchan-delete}, 
where posts 
are withdrawn as users stop contributing to them for a prolonged time.

\paragraph{Problems with premeditated withdrawals: No historical data} 
The above premeditated withdrawal methodologies remove every post from the public view eventually; 
thus, there is {\em no} archived history of user data. However, existence of archival data can be important to not only the system but also the users.
A recent survey~\cite{bauer-2013-postanachronism} shows that users have a keen interest 
in going back to the past social content they 
have uploaded, e.g., for reminiscing old memories. 
Moreover, as social media sites are often perceived as a mirror of the real world, 
reflecting events in the past and how people reacted to them, archiving the past uploaded content has immense historical value;
e.g., US Library of Congress~\cite{libraryOfCongress-blog} is already archiving all uploaded public Twitter data.

Moreover, if a user deletes her post before the predefined time (or viewership) limit on the post,
an adversary can be certain that it is a user-initiated content deletion.
In this case, the current premeditated schemes provide no privacy or deniability to the user.

\paragraph{Our Approach}
Our challenge is to devise a privacy mechanism
that offers protection to user-initiated content deletions (from a persistent onlooker with pervasive access)
without reducing the content's archival value.
We demonstrate how to achieve these contrasting privacy and availability goals 
by systematically withdrawing and resurrecting non-deleted posts from public view.


\section{Problem and Key Idea}

\subsection{System and Adversary Model}\label{sec:threat}

\noindent We model a user-generated data sharing platform (e.g., Twitter) as a public bulletin board where individuals can upload and/or view content. Below we define prominent players and their roles in our setup: 
{\em Platform} is the system,
which maintains the bulletin board (used to upload and view user generated content);
{\em Data Owner} is a user who uploads her posts to the bulletin board.
{\em Adversary} can view the uploaded posts on the bulletin board 
and is constantly in search of posts which have been deleted by their owners 
(possibly to scavenge for the posts that are sensitive to their owners).

In our generic model, all the subscribers (including the adversary) have complete access to the bulletin board 
and can view the posts as they wish. After a data owner decides to delete a post, 
the post will be removed from the bulletin board and will not be visible to anyone. 
We expect the publisher to be honest and assist towards achieving the privacy goal.


Our adversary accesses the bulletin board continuously and takes snapshots at will.
He can determine the deleted posts by comparing the two snapshots taken at different times and pinpointing the posts that existed in the first one but not in the second one (the same strategy used to find deleted tweets in previous studies~\cite{maddock-2015-missingData,
	mondal-2016-longitudinal-exposure}).
The adversary is capable of adding posts and deleting them from the bulletin board; however, it will not be able to delete some other users' posts.
Although the adversary is ultimate in terms of the data access, 
given the manual nature of the task of determining sensitive deletions, 
his goal will be to flag and analyze as few non-deleted posts as possible.
In the real world, an adversary would be actually limited in its capability; consequently, all the privacy guarantees we observe in this work are actually lower bound (Section 8). 
Finally, we expect all aspects of our system and its parameters to be public,
and the adversary to be aware of those.

\subsection{Security Goals}\label{sec:goals}

\noindent Towards our goal to conceal deletions from the adversary 
without significantly affecting the availability, 
we propose the following security properties:

\noindent{\bf Deletion privacy} is the uncertainty of the adversary about a \post{} having been deleted or just temporarily withdrawn by the platform at a given point of time. 
In other words, it is the deniability of deleting a post for the data owner.
As the post remains down for a longer duration, the adversary becomes more certain about its deletion, 
achieving a particular level of privacy is directly related to having a certain \textit{Decision Threshold} 
on the observed down periods for declaring that posts are deleted beyond that point.

\noindent{\bf Platform availability} represents the average availability of a post within a period. The goal is to provide privacy guarantees to users while obtaining high levels of availability.
It is easy to observe that introducing down periods creates a trade-off between privacy and availability. For example, assuming the mean up duration is fixed, as the mean of down distribution increases the availability of the \platform{} will decrease; however, when a post is deleted, it remains unnoticed to the adversary for longer periods due to higher decision thresholds.

It is natural to ask 
why the adversary cannot select his decision threshold independent of down distribution 
(and subsequently availability).
The answer lies in the difficulty of distinguishing sensitive posts from non-sensitive ones.
Sensitivity of a post varies from person to person, and also with life events and time in general, therefore, 
pinpointing sensitive posts for each user is a hard task.
Moreover, there is a huge discrepancy between the content creation and deletion rates on social sites today
(social sites are generating new content at the rate around ten times more than deletions).

This brings us to our third property of adversarial overhead as we expect our
adversary to be concerned with flagging many non-deleted posts (false positives).

\paragraph{Adversarial overhead}is associated with the number of non-deleted posts falsely flagged 
as deleted (false-positives) that the adversary has to investigate along with the detected actual deleted posts (true-positives). We capture it by the precision measure:

\vspace{1mm}
$Precision = \dfrac{True\ Positives}{True\ Positives + False\ Positives}.$
\vspace{2mm}

Towards offering a balanced viewpoint, we also consider the recall measure 
capturing false-negatives (i.e., posts that are flagged as non-deleted but will eventually be deleted):

$Recall = \dfrac{True\ Positives}{True\ Positives + False\ Negatives}.$
\vspace{2mm}

There is a trade-off between privacy and adversarial overhead similar to the trade-off between privacy and availability. 
Ideally, the adversary overhead should be high which implies that the precision should be low.
If the adversary needs to keep its overhead low (less 
false positives), 
it has to provide better privacy (deniability) to its victim by increasing its decision threshold period.

\subsection{Key Idea}

\noindent We plan to provide privacy for a post deletion by intermittently withdrawing the non-deleted posts 
such that
the adversary cannot distinguish between a temporarily withdrawn post and
a permanently deleted post for some long time duration after the deletion.
At its core, our intermittent withdrawal mechanism consists of choosing 
alternating up and down periods of random durations. 
This obviously adversely affects the availability of posts: 
increasing withdrawal time of a post can improve the deletion privacy; 
however, it reduces the overall availability.
Therefore, our key challenge is to determine distributions (and their parameters) for
these intermittent withdrawals such that we achieve a satisfactory level of deletion privacy
without significantly affecting the availability of the posts.

We illustrate our distributions selection process through the following two Straw-man proposals.

\paragraph{Straw-man proposal~I}
As a simple example, consider the degenerate (or fixed-value) distribution
for up and down duration of a post. With $90\%$ availability in mind, 
we consider an alternating series of fixed up period of nine hours and fixed down period of an hour.
Here, every post once withdrawn remains down for a complete hour.
Thus, the adversary cannot flag a post as deleted until it remains down for more than an hour
as any flagging during the first hour down time cannot be better than just randomly flagging
the posts. 
However, the adversary becomes certain about the deletion right after this one hour of down period.
Moreover, if the deletion occurs sometime during the up period of nine hours, 
the adversary can break the privacy immediately.

Although it is possible to increase down time while maintaining the same availability, 
the adversary can simply wait longer before becoming certain about the deletion. 
Larger down time may also not be acceptable to platforms expecting
content to be highly available.

\paragraph{Straw-man proposal~II}
We can replace the above degenerate distribution
by the uniform distribution with mean value of nine hours for the up distribution
and mean value of one hour for the down distribution. 
Here, the deletion can happen anytime during the up duration 
without the adversary becoming certain about the deletion. However,
the problem with the down period remains: with the finite support of the down distribution
(two hours for our example), the adversary will be sure about deletion after two hours. 

\paragraph{Towards \system}
As we do not expect the platform and the users to accurately predict the waiting time
(i.e., decision threshold) for the adversary, 
we propose to use the distributions with infinite support.
Here, the adversary can {\em never} be certain about the deletions; 
but it is easy to see that once the post is deleted, 
the adversary becomes more certain about it as time progresses.

Towards building and analyzing \system, we measure privacy as likelihood ratio in Section~\ref{sec:formal_def},
and find it to be inversely proportional to both hazard rate of the up distribution 
and  complementary cumulative distribution of the down distribution.
We measure availability as the ratio of mean up distribution and
sum of means of both (up and down) distributions.  
In Section~\ref{sec:Design}, we then explore different distributions with infinite support to select
an up and down distribution that offers an excellent trade-off between
deletion privacy, availability and adversarial overhead. 
Finally, in Section~\ref{sec:evaluation},
we evaluate the system for the estimated Twitter dataset.

\subsection{Non-goals}\label{sec:non-goals}

\noindent While solving this complex problem towards achieving privacy, we make some simplifying assumptions.

Firstly, we consider all withdrawn posts to be \textit{equal}, and do not consider the sensitivity of a post's content.
Several other studies~\cite{krishnamurthy-2009-privacyleak,srivastava-2013-privacyLeak,krishnamurthy-2011-disconnect} investigated the sensitivity of posts in general and resulting privacy leaks. 
Those studies provide complementary privacy guarantees and can be used in addition to our approach. 

Secondly, we 
do not take into account correlations between posts, and
instead, assume individual posts to be independent in this first 
proposal for a very difficult problem. Given extremely unpredictable and context-dependent nature of correlations between posts on social sites, considering correlations where they are apparent, will be an interesting future work.

Finally, similar to the usage of salting in password hashing against the dictionary (or rainbow table) attacks, 
our goal is to protect the privacy of withdrawn posts on a \textit{large scale}, and our adversary scavenges through all the withdrawn posts to find as many sensitive deletions as possible. 
We do not aim to protect against a devoted stalker who stalks a particular user or post over a long duration. 
For example, an adversary with prior knowledge of users (e.g., posting patterns) will have an advantage that we do not consider.
Nevertheless,  as compared to the state-of-the-art, 
we aim at increasing the workload of devoted attackers and at delaying the deletion privacy loss
at least by a few arguably important weeks.

\section{Problem Formalization} 
\label{sec:formal_def}


\subsection{Formalized Intermittent Withdrawals}\label{section:formal_def}

\noindent In the proposed system, time is discretized in seconds. We denote by $t_c$ the current time.  
We treat each \post{} independently, 
and therefore, the privacy and availability analyses focus on an individual \post{}.
Let $\create$ denote the creation time of the \post{}. 

The intermittent withdrawal mechanism introduces a disconnection between the real state of a \post{} (deleted or non-deleted) 
and the observed state of the \post{} (publicly visible or withdrawn). 
The real state of the \post{} is  available \textit{only} to the \platform{} and the owner,
while the adversary can only see the observed state of the \post{}.

\begin{figure}[t]
	\centering
	\includegraphics[width=0.48\textwidth]{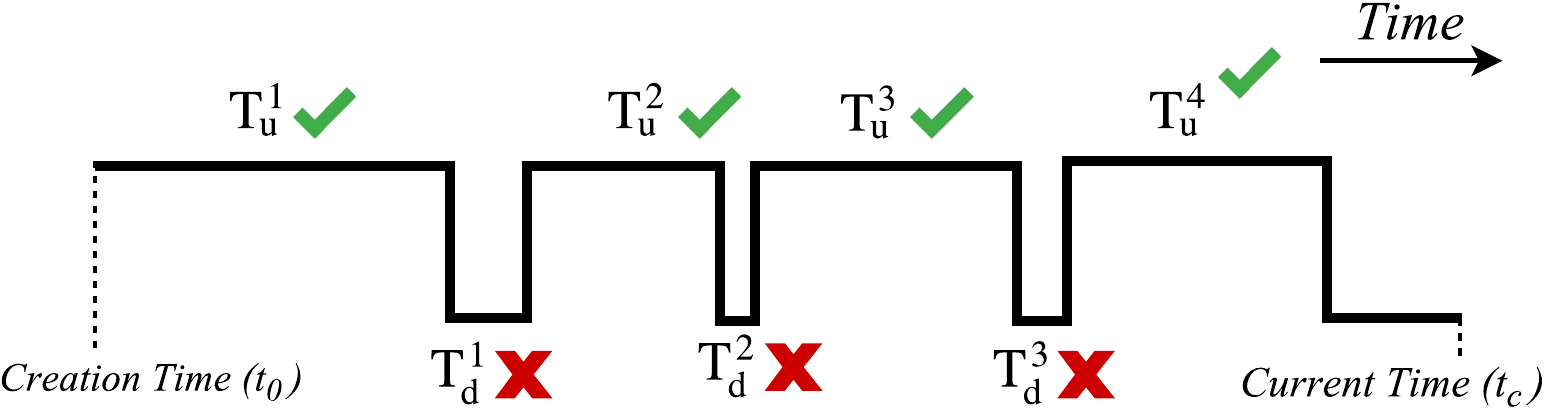}
	\caption{\textbf{Timeline of a \post{}. The \post{} is created at time $\create$, $T_{u}^i$ is the duration of an up phase and $T_{d}^i$ is the duration of a down phase. In  up phases the \post{} is visible to the adversary, in down phases it is not.}}\vspace{-6mm}
	\label{fig:timeline_up_down}
\end{figure}

\noindent \textbf{Real state}: Let $\state{t}$ denote the real state (either non-deleted or deleted) of the \post{} at time $t$. By convention, we say that
$\state{t}=1$ if the \post{} is not deleted at time $t$ and $\state{t}=0$ if the \post{} is deleted.
For example, at creation time $\create$, $\state{\create} = 1$.

We assume that a \post{} cannot be undeleted (or resurrected) and thus can be deleted only once.
Consequently, we define the deletion time $\del>\create$ such that $\state{t}=1$ for all $t\in [\create, \del)$ 
and $\state{t}=0$ for all $t\ge \del$. 
We also assume that $\del$ is not a choice variable of the \platform{} 
and remains unknown to the platform at any time before $\del$.

\noindent \textbf{Observable state}: At any time $t$, by accessing the bulletin board,
the adversary or any user only sees if the \post{} is up (visible) or down (withdrawn). 
Let $\obs{t}$ denote this observable state of the \post{} at time $t\ge \create$. 
By convention, we say that $\obs{t}=1$ if the \post{} is up and $\obs{t}=0$ if the \post{} is down. 

For a \post{}, the \platform{} can decide $\obs{t}$ for all $\create < t  < \del$. In particular, for each \post{}, 
the \platform{} chooses a sequence of positive integer values 
$(T^j_u)_{j\in \mathbb{Z}_+}$ and $(T^j_d)_{j\in \mathbb{Z}_+}$, interpreted as up and down time durations respectively.
The observable state is set as follows.
\begin{small}
	\begin{align}
		\label{eq.Obs} & \!\!\!\!\!\!\!\!\!\textrm{For all } t\in [\create, \del):  \\
		\nonumber & \!\! \obs{t}=1 \textrm{ if, for some } i\ge0, \\
		\nonumber &  \;\;\;\;\; t \in \left[\create + \sum_{j=1}^i T^j_u + \sum_{j=1}^i T^j_d, \create+\sum_{j=1}^{i+1} T^j_u + \sum_{j=1}^i T^j_d\right);\\
		\nonumber & \!\! \obs{t}=0 \textrm{ if, for some } i\ge0, \\
		\nonumber &  \;\;\;\;\; t \in \left[\create + \sum_{j=1}^{i+1} T^j_u + \sum_{j=1}^i T^j_d, \create+\sum_{j=1}^{i+1} T^j_u + \sum_{j=1}^{i+1} T^j_d\right).\\
		\nonumber & \!\!\!\!\!\!\!\!\!\textrm{For all } t\ge \del: \obs{t}=0. 
	\end{align}
\end{small}

\cref{fig:timeline_up_down} illustrates the observable state (from an adversary's point of view) for a \post{} due to the sequences of up and down duration.
As the deletion time $\del$ is 
not known to the platform at any time before $\del$, 
we can assume without loss of generality that 
large sequences $(T^j_u)_{j\in \mathbb{Z}_+}$ and $(T^j_d)_{j\in \mathbb{Z}_+}$ are chosen by the \platform{} at the creation time $\create$. 
As a result, the observable state in Equation~\eqref{eq.Obs} can be intuitively interpreted as follows.
The \post{} is initially up and stays up for a duration $T_u^1$. After the duration $T_u^1$, 
it goes down and stays down for a duration $T_d^1$ before coming up again. 
This process continues indefinitely until the \post{} is deleted by the owner. 
Finally, when a \post{} is deleted, 
it  goes down immediately  even 
if it is in middle of an up duration, and stays down forever. 


Our objective 
is to control the observable state so that it becomes difficult for the adversary 
to be certain about the deletion of a \post{}. In the proposed intermittent withdrawal mechanism,
$(T^j_u)_{j\in \mathbb{Z}_+}$ and $(T^j_d)_{j\in \mathbb{Z}_+}$ are mutually independent i.i.d. sequences of random variables drawn from probability mass functions (PMFs) $f_{T_u}$ and $f_{T_d}$ respectively. 
We define the intermittent withdrawal mechanism as follows:
\begin{definition}[Intermittent withdrawal mechanism]
	We define $\mathcal{M}_{IW}(f_{T_u}, f_{T_d})$ as an algorithm that draws mutually independent  i.i.d. sequences $(T^j_u)_{j\in \mathbb{Z}_+}$ and $(T^j_d)_{j\in \mathbb{Z}_+}$ from $f_{T_u}$ and $f_{T_d}$ respectively, and sets $\obs{t}$ as in Equation~\eqref{eq.Obs}.
\end{definition}
As elaborated later in \cref{sec:Design} and onwards,
We choose parameters PMFs $f_{T_u}$ and $f_{T_d}$ of the $\mathcal{M}_{IW}$ 
to satisfy the contrasting privacy, availability, and adversarial overhead requirements. 
Throughout the analysis, $F_{T_u}$ and $F_{T_d}$ represent the cumulative distribution functions (CDFs),
and $\overline{F_{T_u}}$ and $\overline{F_{T_d}}$ represent the complementary cumulative distribution functions (CCDFs) 
of $f_{T_u}$ and $f_{T_d}$ respectively. We assume that the \platform{} can efficiently sample values from distributions $f_{T_u}$ and $f_{T_d}$,
and that these distributions are known to the adversary. 

Next, we formally analyze our security goals 
in the context of $\mathcal{M}_{IW}(f_{T_u}, f_{T_d})$.

\begin{figure}[b]
	\centering
	\includegraphics[width=0.4\textwidth]{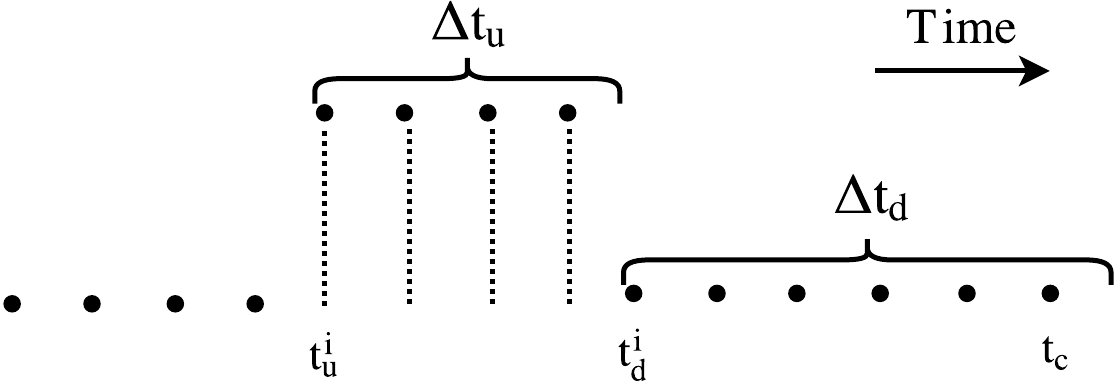}
	\caption{\textbf{Observing the status of a single post from its creation and precisely looking at the last up and down duration, $\upDur$ and $\downDur$ respectively. $t_c$ is the current time, $t_u^i$ denotes the last up toggle and similarly $t_d^i$ is the last down toggle time.}}\vspace{-3mm}
	\label{fig:lastUpDown}
\end{figure}

\subsection{Deletion Privacy}\label{sec:privacy_prop}

\noindent The notion of deletion privacy should quantify the uncertainty of the adversary in distinguishing between 
a \post{} being really deleted by the owner or just in one of its down durations. 
We define this likelihood of adversary detecting an actual deleted \post{}
as the likelihood ratio of the observed sequence of observable states since \post{} creation conditioned on the \post{} being deleted 
or not at the current time $t_c$.

\begin{definition}\label{def.privacy}
	For any time $t_c$, we define the privacy of mechanism $\mathcal{M}_{IW}(f_{T_u}, f_{T_d})$ as a ratio (LR) 
	\begin{align}
		LR = &
		\dfrac
		{\sup_{t\le t_c} Pr( \nobs_{\mathcal{M}_{IW}}([\create, t_c]) \mid \del = t)}
		{\sup_{t> t_c} Pr( \nobs_{\mathcal{M}_{IW}}([\create, t_c]) \mid \del = t )}
		\label{eq:main_eq},
	\end{align}
	where $\nobs_{\mathcal{M}_{IW}}([\create, t_c])$ is the observed state for posts due to ${\mathcal{M}_{IW}}$
	in the interval $[\create, t_c]$.
\end{definition}

The above ratio is the classical \textit{likelihood ratio} (LR) statistic \cite{Casella02a} for the test to determine if the \post{} was deleted or not, i.e., the test with null hypothesis $H_0: \{\state{t_c} = 1\}$ (equivalent to $\{\del > t_c\}$) and alternative hypothesis $H_1: \{\state{t_c} = 0\}$ (equivalent to $\{\del \le t_c\}$). 
It is known that likelihood ratio tests have good properties and are often the most powerful tests that the adversary can do to determine if the \post{} was deleted \cite{Casella02a}. 
Hence, limiting this likelihood ratio is the best way of limiting the possibility for the adversary of accurately testing if the \post{} was deleted or not. 
Increase in the LR value for a \post{} denotes increase in certainty of the adversary about a \post{} deletion;
in short, lesser value of LR denotes better privacy. 
Since the adversary knows the up and down time distributions 
it can compute the likelihood ratio of the deletion privacy. Our definition of deletion privacy parallels with the definition of differential privacy~\cite{Dwork-2006-dp}, however there is subtle difference between them (Appendix~\ref{appendix:diff}).
In this paper we analyze the privacy using the Frequentist approach. For interested readers in the Bayesian analysis we refer to Section 4 of~\cite{lethePets2019}.


\paragraph{Deletion Privacy for the Intermittent Withdrawal Mechanism}

As deletion privacy (or LR value) depends on $\nobs([\create, t_c])$ 
(i.e., the sequence of observable states chosen by the \platform{}) and consequently 
on the distributions $f_{T_u}$ and $f_{T_d}$, 
we need to quantify this dependency to understand the deletion privacy offered by intermittent withdrawal mechanism.

In our intermittent withdrawal mechanism, up and down durations are drawn i.i.d. until the \post{} is deleted. 
Therefore, the probability of the sequence is the product of probability of observing each duration which is the same regardless of 
if the \post{} was deleted or not except for the last up and down durations;
one of the last up and down durations could be cut by the deletion. 
As a result, the ratio $LR$ depends only on the last up and down durations. 
We denote last up and down duration by $\upDur$ and $\downDur$ respectively 
and by extension by $\obs{\upDur,\downDur}$ the observed state in those times (see illustration on \cref{fig:lastUpDown}). 
Then the likelihood ratio on the lhs of~\eqref{eq:main_eq} can be simplified as
\begin{equation}
	LR = \dfrac
	{\sup_{t\le t_c}  Pr( \obs{\upDur,\downDur} \mid \del = t)}
	{\sup_{t > t_c}  Pr( \obs{\upDur,\downDur} \mid \del = t)}.
	\label{eq:main_last_updown}
\end{equation}

Now we compute the numerator and denominator separately. The denominator is simply the likelihood of observing $\obs{\upDur,\downDur}$ if the \post{} was not yet deleted at time $t_c$ (i.e., $\state{t_c} = 1$), which is
\begin{equation}
	Pr( \obs{\upDur,\downDur} \mid \state{t_c} = 1) = f_{T_u} (\upDur) \cdot \ccdfDo{\downDur - 1}. \label{eq:main_denom_res}
\end{equation} 
As the \post{} has not been deleted at time $t_c$ 
(i.e., $\state{t_c} = 1$), 
the probability of observing $\upDur$ is $\fUp{\upDur}$. Moreover, since the \post{} is in middle of a down period 
the probability of observing $\downDur$ is $Pr(T_d^i \ge \downDur) = \ccdfDo{\downDur-1}$. 

For the numerator, we compute the probability of $\obs{\upDur,\downDur}$ conditioned on the deletion time being $t$ for each $t$ between the last toggle $t_d^i$ and the current time $t_c$ and take the maximum of those probabilities. (It is {\em not} necessary to consider earlier deletion times since the probability of $\obs{\upDur,\downDur}$ would then be zero.) We treat separately the case where $\del=t_d^i$ which corresponds to a deletion happening during (or at the end of) an up period and the cases $\del \in (t_d^i, t_c]$ which correspond to a deletion happening during a down period. 
In the second case, for $t \in (t_d^i, t_c]$, we have 
$$
Pr( \obs{\upDur,\downDur} \mid \del = t) =  \fUp{\upDur} \cdot \ccdfDo{t - t^i_d-1},
$$
which is maximized for $t=t^i_d+1$ where $\ccdfDo{t - t^i_d-1} = \ccdfDo{0} = 1$. In the case where $\del=t_d^i$, then the last up period could have been either of exactly $\upDur$ or of more, hence
$$
Pr( \obs{\upDur,\downDur} \mid \del = t^i_d) =  \ccdfUp{\upDur} + \fUp{\upDur}.
$$
Since $\ccdfUp{\upDur}\ge 0$, we conclude that 
\begin{equation}
	\sup_{t\le t_c}  Pr( \obs{\upDur,\downDur} \mid \del = t) = \ccdfUp{\upDur} + \fUp{\upDur}. \label{eq:main_num_res}
\end{equation}

Substituting \eqref{eq:main_denom_res} and \eqref{eq:main_num_res} into \eqref{eq:main_last_updown}, we get the final expression of the likelihood ratio: 
\begin{align}
	LR &
	=\left({\frac{\ccdfUp{\upDur}}{\fUp{\upDur}} + 1 }\right)
	\cdot \frac{1}{\ccdfDo{\downDur - 1}}. \label{eq:main_res}
\end{align}

Equation~\eqref{eq:main_res} 
captures the relation between the LR (i.e., deletion privacy) and the choice of up and down time distributions: 
(i) the LR is (almost) inversely proportional to the \textit{hazard rate} $\fUp{\upDur} / \ccdfUp{\upDur}$ of the up distribution; and (ii) the LR is inversely proportional to the \textit{CCDF} $\ccdfDo{\downDur - 1}$  of the down distribution. We need to optimize for these two functions while choosing up and down time distributions for controlling privacy guarantee of $\mathcal{M}_{IW}$.

\subsection{Availability property}\label{sec:avail_prop}

\noindent The intermittent withdrawal mechanism provides deletion privacy at the cost of reducing availability of the \post{}. The \post{} is not visible to the adversary as well as any benign observer during the down periods. 
Intuitively, the availability of a \post{} is simply the fraction of time the \post{} is visible to an observer. Formally, for mechanism $\mathcal{M}_{IW}(f_{T_u}, f_{T_d})$ the availability is:
\vspace{-1mm}
\begin{equation}
	Availability = \dfrac
	{\mu_{f_{T_u}}}
	{\mu_{f_{T_u}} + \mu_{f_{T_d}}},
	\label{eq:main_avail}
	\vspace{-1mm}
\end{equation}
where $\mu_{f_{T_u}}$ is the mean of the up time distribution $f_{T_u}$ and $\mu_{f_{T_u}}$ is the mean of the down time distribution $f_{T_d}$. 

The LR~\eqref{eq:main_res} and availability~\eqref{eq:main_avail}, both are functions of the up and down time distributions and thus are correlated. For instance, when \posts{} in the archive are always down (e.g., $f_{T_u}$ is a finite distribution and $f_{T_d}$ is a distribution with infinite mean), the archive has zero availability and perfect privacy (the LR value is 1). On the other hand, when \posts{} in the archive are always up (e.g., $f_{T_d}$ is a uniform distribution with mean $0$), the archive has perfect availability of 1 and no privacy (LR value is $\infty$). 
In non-extreme cases, the relationship of availability and privacy is more intricate 
and depends on specific choices of up and down distributions.
We explore this trade-off empirically in \cref{sec:evaluation}. 
	
\section{\system\ Design}\label{sec:Design}

\noindent 
We parameterized the security guarantees in section~\ref{sec:formal_def}, but
we still need to determine exact specifications for these parameters to effectively control the guarantees. 
The required parameters include the mean up (down) times for the up (down) distributions 
as well as choices of PMFs for those distributions. 
The key design challenge for \system{} is: \textit{How to choose suitable parameters for \system{}  to give good availability and privacy guarantees?} Here, we resolve this design challenge empirically.

\subsection{Choosing up/down distribution mean values to control availability}

\noindent Availability of \system{}, the average fraction of up time, depends upon the mean 
for up and down distributions (Equation~\eqref{eq:main_avail}). 
While choosing mean values of up and down time distributions,
the platform operator needs to decide upon the required availability of the \platform{}. 
From a practical perspective, we envision that the \platform{} would need the availability to be around 90\%.

The absolute value of the down time is also interesting from a usability viewpoint: 
Hypothetically if an operator expecting 90\% availability sets the mean down time as one year and mean up time as nine years, 
a particular post will be hidden on average for one year. 
However, a year of down time on average is unacceptable in many real-world scenarios: 
the users may leave the system if the non-deleted content is not available for such large durations. 
Therefore, unless otherwise stated, we set mean for down time distributions as one hour  and mean for up time distributions as nine hours. 


\subsection{Choosing up/down distribution PMFs to control  deletion privacy}\label{sec:ultimate_analysis}

\noindent The \platform{} operator needs to control the deletion privacy guarantee of \system{} via setting some suitable choices for up and down time distributions (i.e., their PMFs). 
Her aim is to 
minimize the LR value. 

\begin{figure}[t]
	\centering
	\includegraphics[width=0.95\columnwidth]{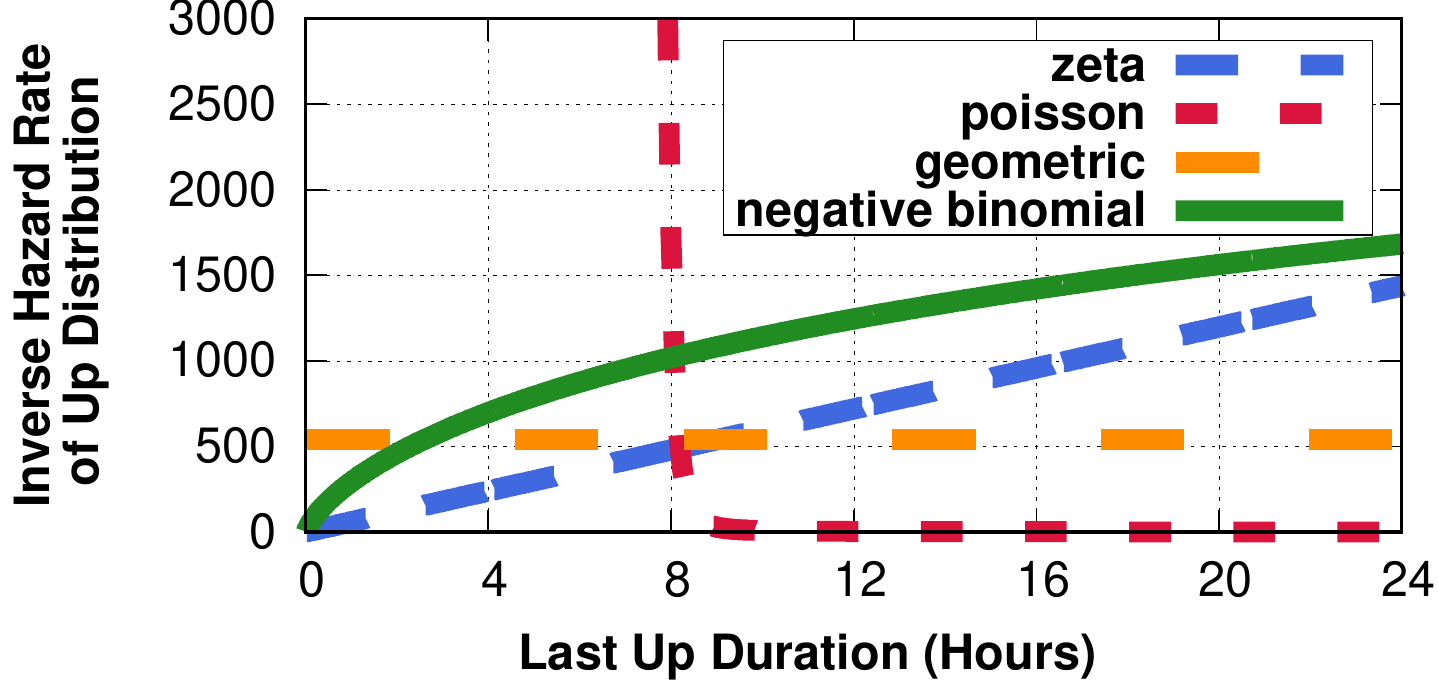}
	\vspace{-2ex}
	\caption{\textbf{Variation of inverse hazard rate with time for four choices of up time distributions (with same mean). Increase in inverse hazard rate signifies increase in LR value.}}\vspace{-3ex}
	\label{fig:ultimate_up_diff_dist}
\end{figure}

\paragraph{Geometric distribution is a suitable choice for up time distribution} 
Recall that the value of LR, 
is inversely proportional to the hazard rate for the up time distribution (\cref{eq:main_res})
at the last up duration. 
To select the up times distribution, we considered a wide range of distributions varying in their main characteristics and we present here four distributions
with infinite support and the same mean of nine hours---zeta, poisson, geometric and negative binomial~\cite{walck2007handbook}---that illustrate the main rationales behind our choice.
\cref{fig:ultimate_up_diff_dist} shows the inverse hazard rate for these four choices
of up time distributions for different values of last up durations (ranging up to 24 hours). The trends remain similar for longer time durations.
Note that, negative binomial distribution requires a parameter called the shape parameter or $n$, which is set to 0.15 in  \cref{fig:ultimate_up_diff_dist} for demonstration. The take away in this figure remains the same for other values of $n$. 
The key observation is that only the memoryless geometric distribution has a constant inverse hazard rate for different last up durations. 
If we take geometric distribution as our up time distribution function, 
\textit{any} value of last up duration will have the same effect on the value of LR, i.e., the value of LR will not be affected even when a deletion happens in middle of an up duration (and effectively cut short the original up duration).

However, this is not the case for other distributions---their inverse hazard rate changes with the value of last up duration. Thus, aside from geometric distribution, any other choice of up time distributions poses two problems: (i) the inverse hazard rate (and consequently LR value) would be very high at some point for the last up duration, as evident from ~\cref{fig:ultimate_up_diff_dist} and (ii) if a post is deleted in the middle of last up duration the LR value will change for that post (since deletion effectively changes the original value of last up duration) compare to the case of no deletion. This phenomenon might provide additional hint to the attacker. 
Thus we strongly prescribe to use geometric distribution as a suitable choice of up time distribution. 

We note that our choice is conservative---for other distributions, there will be instances  where inverse hazard rate 
(and subsequently the LR value) is lower compare to geometric distribution (see \cref{fig:ultimate_up_diff_dist}). 
However, we prefer predictability in the inverse hazard rate of geometric distribution (thus value of LR) for a deployment. 

\begin{figure}[t]
	\centering
	\includegraphics[width=0.95\columnwidth]{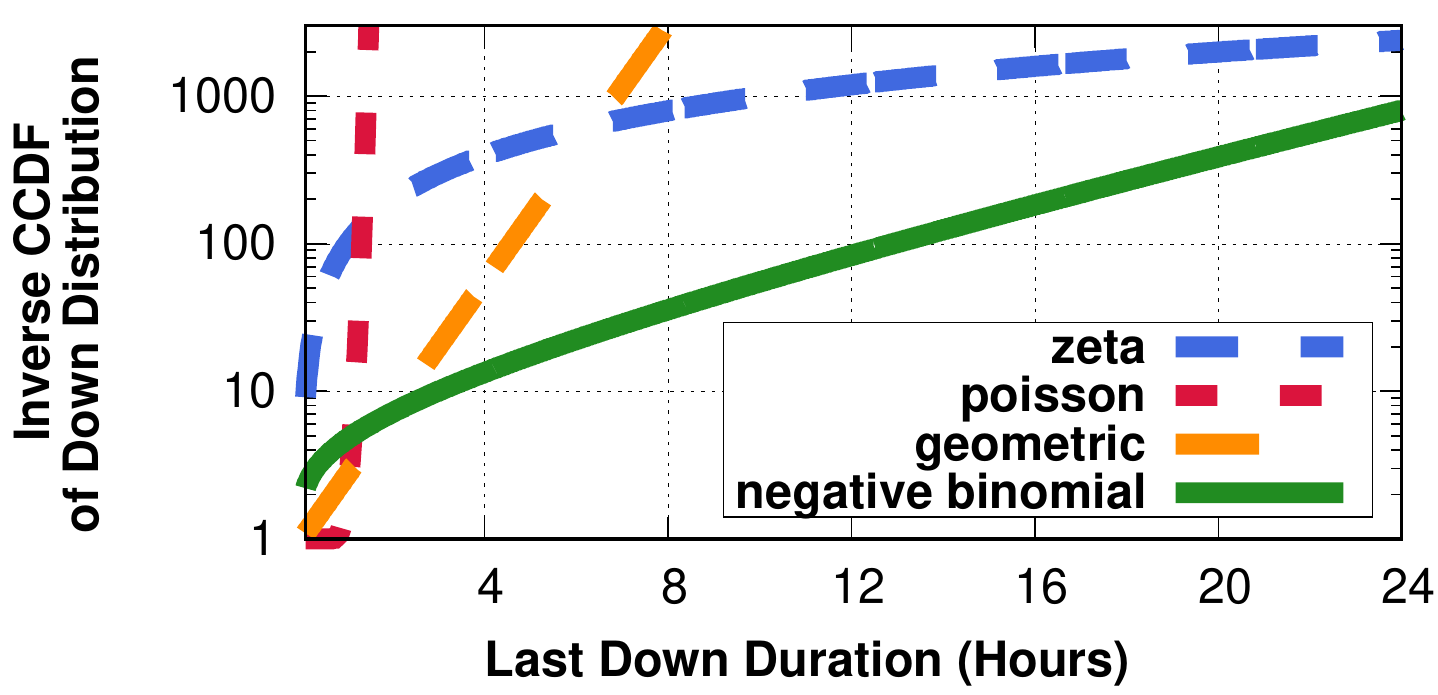}
	\vspace{-2ex}
	\caption{\textbf{Variation of inverse of CCDF values in log scaled (proportional to value of LR) for down distribution with last down duration for four choices of down distributions.}}\vspace{-3mm}
	\label{fig:ultimate_down_diff_dist}
\end{figure}

\paragraph{Negative binomial distribution is a suitable choice for down time distribution} 
Similar to up time distribution analysis, 
we have experimented with a wide range of distributions for down times. 
Recall that the LR value, 
is proportional to the inverse CCDF of a given down time distribution (\cref{eq:main_res})
\cref{fig:ultimate_down_diff_dist} presents the inverse of CCDF of down time distribution 
in log scale for different values of last down time duration 
(ranging up to 24 hours) for our four representative choices---zeta, geometric, Poisson and negative binomial~\cite{walck2007handbook} (each with a mean of one hour). The trends remain similar for longer time durations. 
We first observe that for a small down duration,
the Poisson distribution has the lowest inverse CCDF value (thus lowest LR). 
However, at the mean down duration, 
the value quickly jumps and becomes the highest amongst the different distributions tested. 
The reason is that most values in the Poisson distribution are concentrated around the mean. Hence, before the mean, the CCDF is close to 1
but quickly after the mean it becomes close to zero
(intuitively, for Poisson distribution there is a negligible chance that a non-deleted post observes a down time much larger than the mean; thus observing one gives a very strong signal to the attacker). 

Similarly, any other distribution with value concentrated around a mode would suffer the same limitation and it is preferable to select a distribution with a decreasing PMF such as geometric, zeta or negative binomial. 
Amongst those three, geometric has lowest LR for small down time durations, but it increases rapidly for large down time durations. 
Comparatively, zeta has higher LR for small down time durations and smaller values for large down time duration. This difference is because the geometric distribution has a light tail and its PMF decreases faster whereas the zeta distribution has a heavy tail and therefore assigns higher weights to very large values--hence observing even a very large value has a non-negligible probability to happen under no deletion if the down time distribution is zeta. 
Finally, the key observation from \cref{fig:ultimate_down_diff_dist} is that the inverse CCDF value of negative binomial distribution provides a balance between these two patterns and thus presents itself as a nice choice for down time distribution.

However, there is a challenge while using negative binomial distribution: 
it takes another parameter (in addition to mean down time), called the shape parameter and denoted ``$n$''. 
In \cref{fig:ultimate_down_diff_dist}, $n$ is set to 0.15 for demonstrating trends, 
but a practical deployment of \system{} requires a systematic guideline for setting $ n$. 
Specifically, if the \platform{} operator can have an estimate $\decisionThresEst$ for adversary's decision threshold, 
then it can choose $n$ such that the value of LR is lowest for decision threshold $\decisionThresEst$. 
The \platform{} operator may even base $\decisionThresEst$  on user perception,
e.g., operator decides that it is ok, if an adversary finds out deletion of a post after six months or more.

As evident in \cref{fig:ultimate_down_diff_dist}, zeta distribution will outperform negative binomial distribution at some point in time. However, we claim that for all the decision thresholds that we have considered (even years), 
there exists a shape parameter for the negative binomial distribution 
that provides lower LR value for that threshold compared to zeta distribution.
On the other hand,
if the \platform{} cannot come up with any reasonable $\decisionThresEst$ it might use zeta distribution, 
since eventually it will perform better than negative binomial distribution; 
however, this comes at the cost of lowering privacy, i.e., increased LR value, for some period of time.
In general, we expect the platform operators, based on their experience, to estimate the range of decision threshold $\decisionThresEst$ values reasonably well.

We discuss a procedure to calculate the value of the shape parameter ($n$) of negative binomial distribution (given the mean down time and the adversary's decision threshold) and its effect on the LR value in \cref{section:shape_param_disscussion}.

\subsection{\system\ Algorithm}

\noindent\textit{Input:} 
\platform{} availability percentage, mean down time,
adversary's decision threshold.

\noindent\textit{Algorithm:} 
\begin{compactenum}
	\item Acquire the mean up time based on the provided mean down time and availability values.
	\item Obtain the shape parameter using the derivative procedure based on Equation~\eqref{eq:driv} using 
	the mean down time and decision threshold from input.
	\item Initialize the up and down distributions by passing the mean up and down times along with the shape parameter for the down distribution.
	\item Upon a  post creation, set the real state of the post to $1$ 
	and instantiate the first up period from the up distribution. Set observable state of the post to $1$.
	\item Upon a toggle signal for a post, if the post was in a up period instantiate a down period from the down distribution and set the observable state to zero; Otherwise instantiate an up period from the up distribution and set the observable state to one.  
	\item Upon a deletion request for a post from the owner, 
	set the real and observable state to zero and remove the post from the active set (i.e. posts that toggle).
\end{compactenum}

\vspace{1mm}
These steps provide a \platform{} operator the basic algorithm to run \system{}. However, from a system design point of view a relevant question is---how to efficiently implement these steps? For example, a simple but inefficient (not scalable) implementation for the \platform{} is to just 
assign one process per post to 
track the
observable state for that post (which is toggled due to \system{}). We find that pre-computing future up and down durations and updating them lazily results in efficient \system{} implementation. 
We direct interested readers to~\cref{implementation} for an efficient \system{}
implementation sketch.


\section{Evaluation of \system}\label{sec:evaluation}


\noindent We evaluate the usefulness of \system{} by answering a key question: In practice, how hard is it for an adversary to detect deleted posts in presence of \system~ (adversarial overhead for identifying deleted posts)?

The \posts{}, which are deleted by the users, will be in a down period for an infinite time. Thus, the down period of such \posts{} will at some point exceed the adversarially chosen decision threshold $\decisionThres$ (associated with the LR values) and be flagged by the adversary.  These deleted \posts{}, once correctly flagged by an adversary, constitute the true-positives $TP_\decisionThres$.
Conversely, when a down period $T_d^i$ for some non-deleted \posts{} 
exceed the decision threshold, 
these falsely flagged \posts{} constitute the false positives $FP_\decisionThres$.
On the other hand, the posts that are flagged as non-deleted but will eventually be deleted will be the false negatives $FN_\decisionThres$.
%

Thus, for a decision threshold $\decisionThres$  set by our adversary,  
if his strategy gives the $TP_\decisionThres$, $FP_\decisionThres$  and $FN_\decisionThres$, we measure the adversarial overhead as 
the precision $Precision_\decisionThres = \frac{TP_\decisionThres}{TP_\decisionThres + FP_\decisionThres}$ and the recall $Recall_\decisionThres = \frac{TP_\decisionThres}{TP_\decisionThres + FN_\decisionThres}$.



To evaluate usefulness of \system{} we empirically explore the relation between adversarial precision, availability and decision threshold set by the adversary.

\noindent \textbf{Data Collection:} 
Today, such an intermittent withdrawal mechanism does not exist in the domain of social media and archives. 
To evaluate the feasibility and performance of \system{}, we take Twitter data as a good model platform.
To that end, we need numbers for non-deleted and deleted posts on Twitter, and the rate of deletion and new tweets addition in Twitter.

Using reports such as~\cite{guardian-TwitterSize, nytimes-TwitterSize}, we estimate that there are one trillion non-deleted tweets in the Twitter platform as of 2015. To determine the rates of deletion/addition of tweets, we resort to the 1\% random sample provided by Twitter~\cite{Twitterapi}. Specifically, we collected 1\% random sample for 18 months (from October 2015 to April 2017). In our 1\% random sample, daily on average, 3.2 million tweets are created, i.e. in the whole Twitter 320 million new tweets are created daily. Further, the 1\% sample also provides us deletion notices;
using those notices we determine how many of archived tweets are deleted daily~\cite{almuhimedi-2013-deleteTweets}. We found that on average around 1 million tweets are deleted daily from 1\% sample. So daily, on average 100 million tweets are deleted from the whole Twitter archive.
Thus, the ratio between the volume of deleted and non-deleted tweets in the 
Twitter platform is approximately 0.01\%.
As time passes, this ratio will become smaller (assuming deletion volume will not change too much).
Finally, daily 220 million non-deleted tweets are added to the archive.

\noindent \textbf{Experimental setup:} For our experiment, we set 1 day as our time unit and pick three system availabilities to experiment---85\%, 90\% and 95\%, all with the mean down time of one hour. Consequently, for 85\%, 90\% and 95\% availability the mean up times are respectively 5.7, 9 and 19 hours. Next, we set the up and down time distributions as geometric and negative binomial respectively (as discussed in \cref{sec:Design}). We use \cref{table:recall_n} to set the shape parameter $n$ for our negative binomial distribution.

To make the \system{} simulation feasible with our available resources, we scale down the absolute numbers of deleted/non-deleted tweets to 0.01\% of their original values. In other words, we simulate \system{} on a scaled down version of Twitter (our archival \platform{}).  We consider that our \platform{} contains 100 million non-deleted tweets (0.01\% of 1 trillion) already archived in the \platform{}
Moreover, 32k tweets are created each day and 10k tweets are deleted (thus adding 22k non-deleted tweets each day) in our \platform{}.

\begin{figure}[t]
	\centering
	\includegraphics[width=0.95\columnwidth]{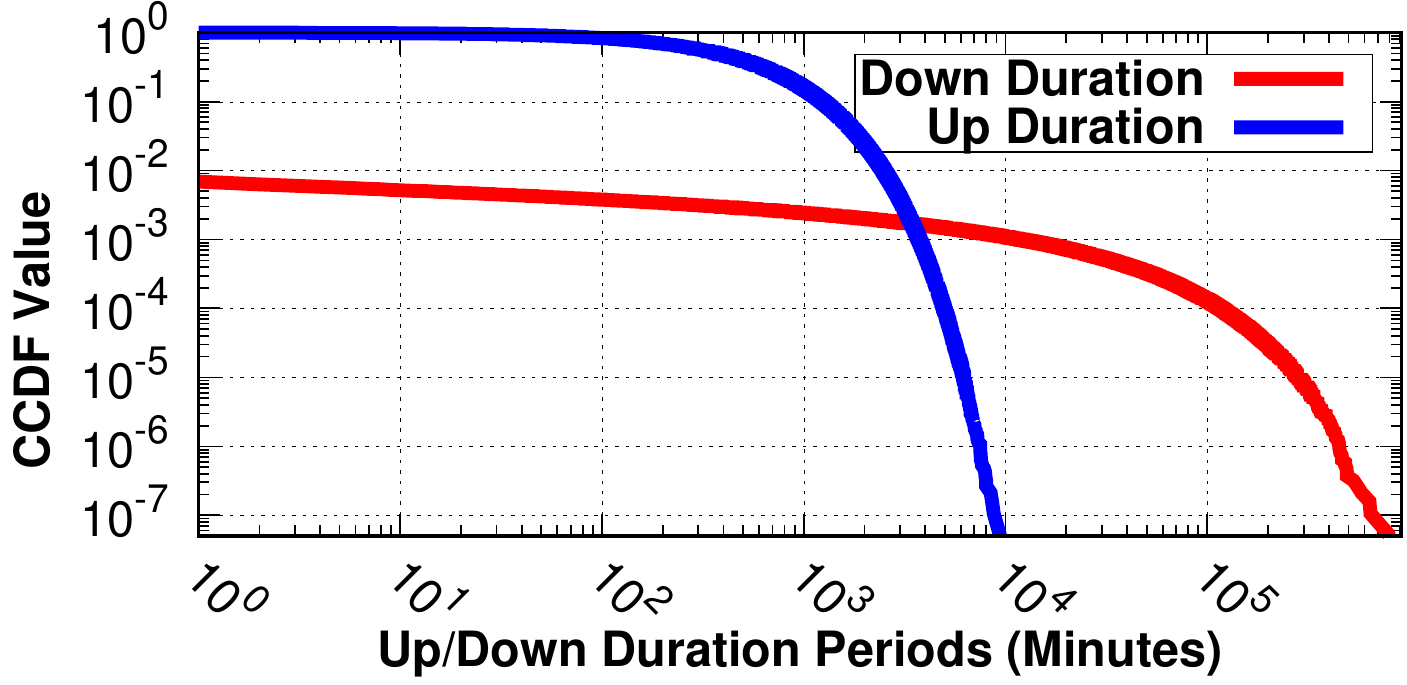}
	\vspace{-2ex}
	\caption{\textbf{CCDF value of up and down durations. The up distribution is a geometric distribution with the mean of 9 hours. The down distribution is a negative binomial distribution with the mean of 1 hour. }}
	\vspace{-3mm}
	\label{fig:ccdf_up_down_durations}
	
\end{figure}

\noindent \textbf{Experimental methodology:} 
For the evaluation of \system{} we take the Frequentist design explained in~\cref{sec:privacy_prop,sec:Design}.
We note that in order to simulate \system{} we don't need the exact timestamps for each \post{} creation and deletion. \system{} is applied to the \posts{} as if all of them were created on the first day of experiment. 
We take 1 day as our time unit and for our simulation, we assume that creation and deletion notifications are received in batch in every time unit. 
We continue this experiment for 10 years (considering creation and deletion of tweets each day). 
\cref{fig:ccdf_up_down_durations} presents the CCDF value of the up and down durations for the chosen distributions at the 90\% availability. More than 99\% of the down durations are less than or equal to one minute. The mean up duration in \cref{fig:ccdf_up_down_durations} is 9 hours and more than 90\% of the up durations are longer than 3 hours.

Leveraging our aforementioned experimental set-up we simulate \system{} and measure adversarial overhead (i.e. precision and recall) at different decision thresholds.
In our set up the true positive for the adversary is simply: number of daily deletions $\times$ (experiment duration - decision threshold). The false positives for our adversary, on the other hand, are non-deleted tweets that get flagged based on the adversary's decision threshold. Further, we note that our adversary might decide to flag the false positives either once or multiple times (i.e., remove flag from a tweet when the tweet is resurrected after a long time and again flag it later). We consider these two scenarios separately.


\begin{figure}[t]
	\centering
	\includegraphics[width=0.95\columnwidth]{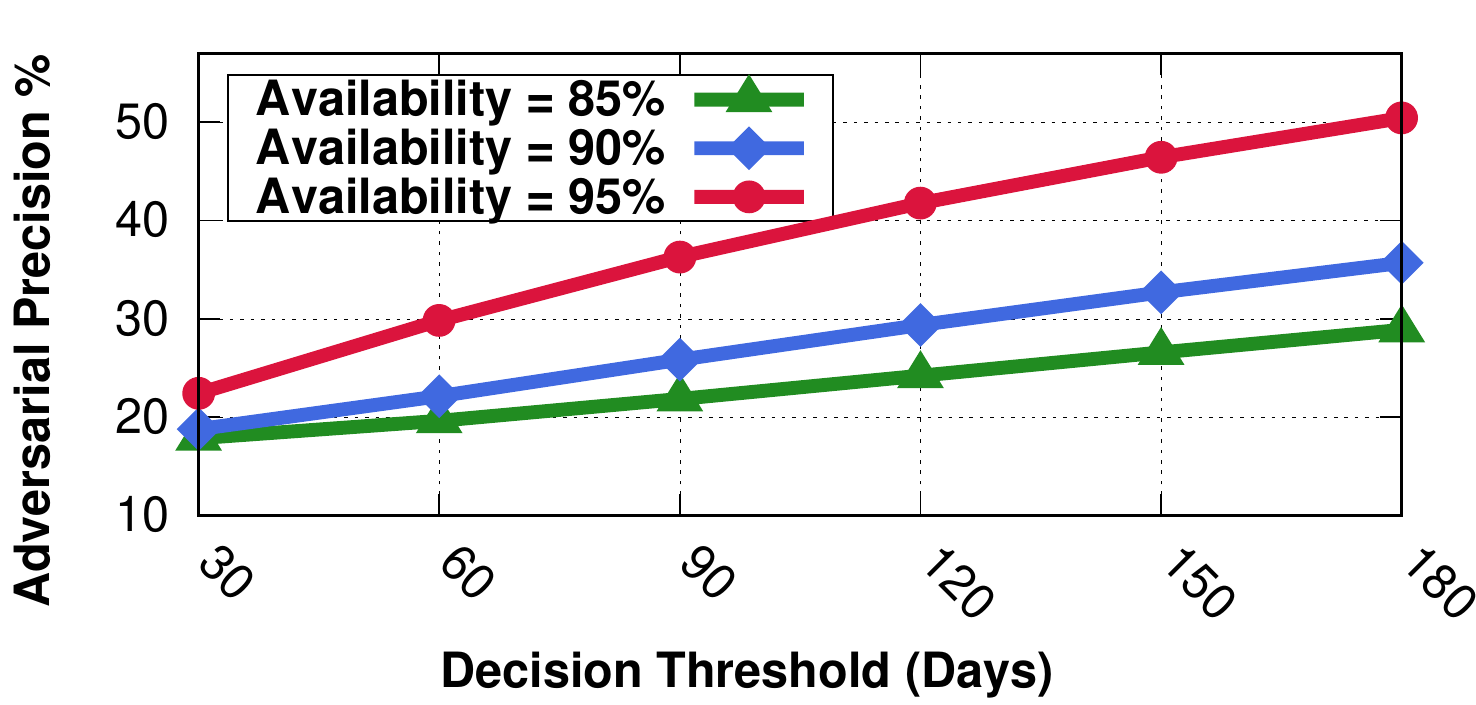}
	\vspace{-2ex}
	\caption{\textbf{Variation of adversarial precision against decision threshold periods for different availability values in flag-once scenario. In this scenario, each tweet can be \textit{flagged only once}.}}
	\vspace{-3mm}
	\label{fig:precision_once}
	
\end{figure}

\noindent \textbf{{Adversary investigates a flagged tweet only once:}} In this scenario, if a non-deleted post gets flagged the adversary will investigate it and after its investigation, it will remove that tweet from his consideration.
Thus, the adversary will not consider the post again in the future, even though the post is visible again. 
We call this scenario \textit{flag-once}.   \cref{fig:precision_once} is showing the variation of adversarial precision for different decision thresholds in X-axis. As the decision threshold increases the adversary's confidence about a tweet being deleted also increases which result in higher precision values. Note that even for 85\% and 90\% \platform{} availability the adversarial precision is around (or less than) 35\% even when the decision threshold is as high as six months or 180 days
, i.e., due to \system{} a deletion will go unnoticed for as long as six months.

\begin{figure}[bt]
	\centering
	\includegraphics[width=0.95\columnwidth]{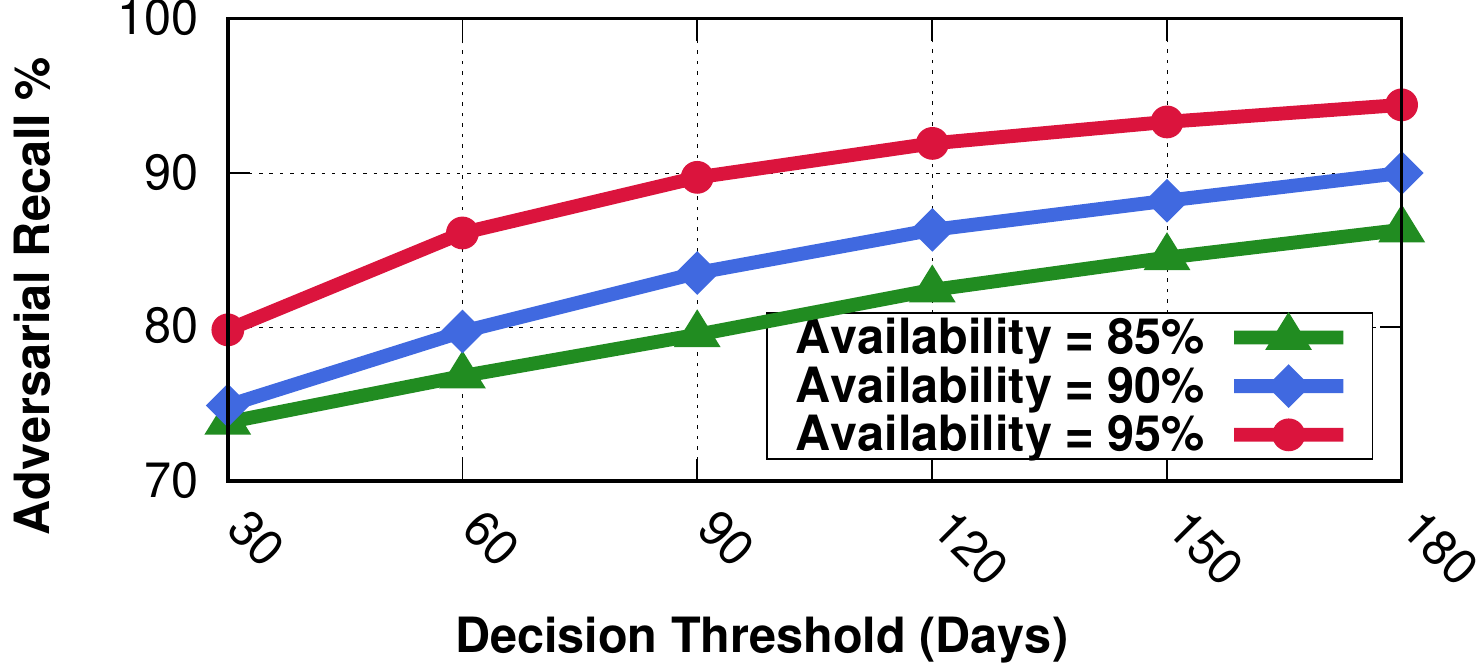}
	\vspace{-2ex}
	\caption{\textbf{Variation of adversarial recall against decision threshold periods for different availability values in flag-once scenario.}}
	\vspace{-3mm}
	\label{fig:recall_once}
	
\end{figure}

This scenario checks a flagged post only once and will not consider it later again. Thus, it is possible that a non-deleted post flagged at time $t$ will actually be deleted at a time later than $t$. So some posts might be deleted but not considered by the adversary, introducing false negatives. \cref{fig:recall_once} shows the variation of adversarial recall of deleted posts for different decision thresholds in X-axis. We make two observations. First, the adversary's recall increases with decision threshold. This is because, with increasing threshold, tweets that are not deleted at time $t$ (but deleted later) will have more time to become visible (not getting flagged) before their actual time of deletion. Second, the recall increases with system availability. The reason is that the number of down periods  decreases with increasing system availabilities and thus it is less likely to obtain  larger down periods to flag tweets. This 
results in higher recall.

\begin{figure}[t]
	\centering
	\includegraphics[width=0.95\columnwidth]{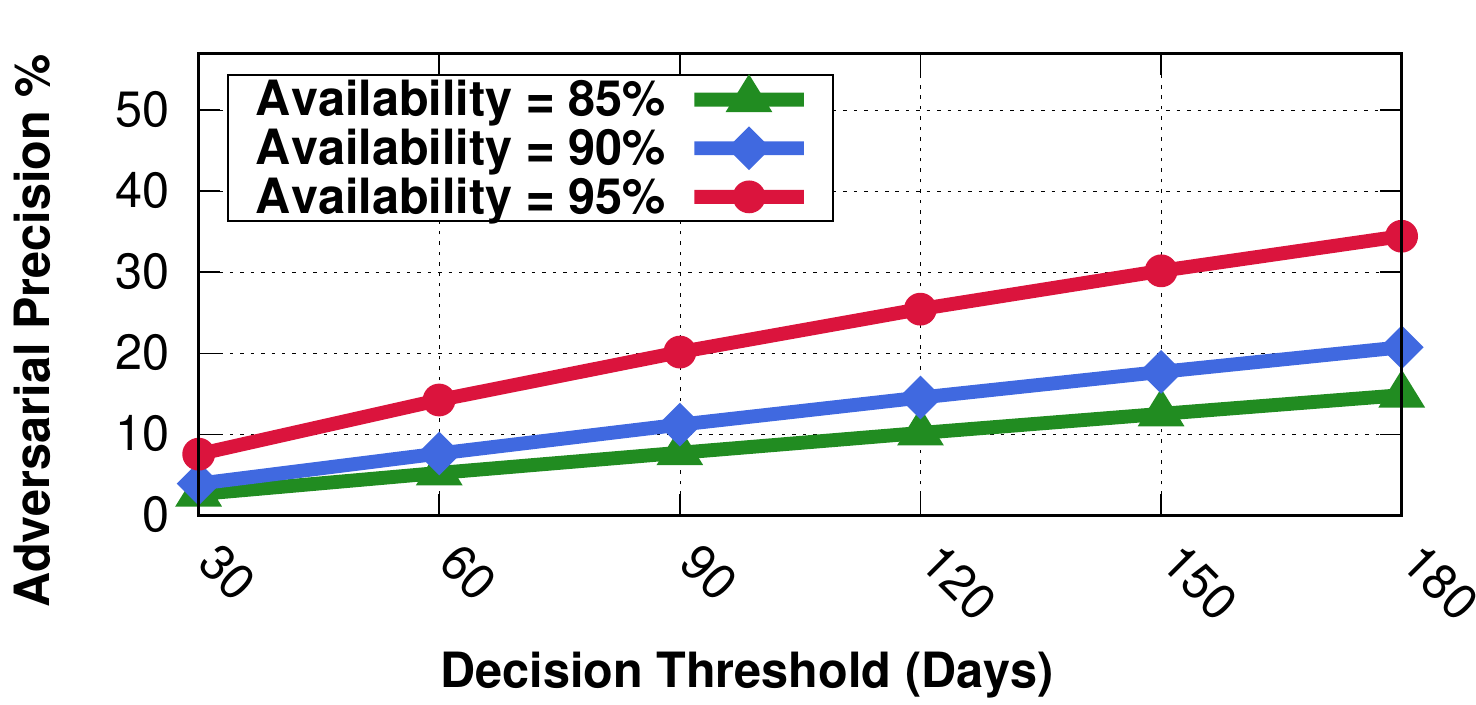}
	\vspace{-2ex}
	\caption{\textbf{Variation of adversarial precision against decision threshold periods for different availability values in flag-multi scenario. In this scenario a tweet can be falsely \textit{flagged multiple times}. }}
	\vspace{-4mm}
	\label{fig:precision_multi}
\end{figure}

\noindent \textbf{{Adversary investigates a flagged tweet multiple times:}}
This scenario is opposite of the previous one in the sense that once a non-deleted tweet has been flagged and investigated it will return to the set of non-deleted. We call this scenario \textit{flag-multi}. The rationality behind this scenario is: it is true that the falsely flagged tweets are not deleted at the current time, but they might be at a future point in time, since sensitivity changes with time and life events. Thus the adversary would also like to take into consideration the real deletion of false positive tweets. \cref{fig:precision_multi} shows the adversarial precision with varying decision thresholds. Compared to the scenario in \cref{fig:precision_once} the adversary has a lower precision for different thresholds for all values of \platform{} availability. The reason is, in this case, a tweet can be flagged multiple times and result in higher false positives. Specifically, in \cref{fig:precision_multi}, for the case of 90\% availability, \system{} keeps adversarial precision around 20\% even when the adversary's decision threshold is as high as 6 months. 

In this scenario if a post is flagged it can again be considered for investigation. Since, a deleted post will remain in a down period forever, the adversary will flag it as soon as the decision threshold is over. Thus, all the deletions will be identified eventually. Consequently, in this case there are no type II errors (false negatives) and recall will always be 100\%.

\noindent \textbf{Overhead of investigating falsely flagged tweets:} Finally, we address one aspect of \system{} that we did not consider so far: the astronomical number of falsely flagged tweets that an adversary has to investigate (i.e., extra work) in either of these scenarios.  \cref{table:false_positives} presents the raw number of non-deleted tweets falsely flagged (i.e. false positives) for both of the aforementioned scenarios. In the worst case, the adversary falsely flagged 13 trillion tweets in the flag-multi scenario when the availability and decision threshold are respectively 85\% and 30 days. As \cref{table:false_positives} shows, even in the best case, with 95\% availability and 180 day decision threshold in the flag-once scenario, the adversary needs to investigate 340 billion falsely flagged tweets.

We have also considered one extreme case---setting the \platform{} availability to 99\% (results not shown), i.e., setting the mean down and up time respectively to 1 hour and 99 hours. Although the precision, in that case, is higher compared to the ones in Figure \ref{fig:precision_once} and \ref{fig:precision_multi}, we found that even with 99\% availability, in the best case (decision threshold 6 months, flag-once scenario) the adversary still needs to investigate 70 billion falsely flagged tweets. In short, We emphasize that the number of falsely flagged tweets is astronomical, and without incurring very high infrastructural cost an adversary can not support such investigation. Thus, much higher decision thresholds are needed for the adversary.

Note that, if an adversary targets a subset of all users, then precision/recall values for both scenarios remain the same and it will only proportionately effect actual number of falsely flagged tweets mentioned in \cref{table:false_positives}. For example, if the adversary is targeting 0.1\% of all the users then number of falsely flagged tweets in \cref{table:false_positives} will be in billions instead of trillions.
Furthermore, as the number of users decreases, the prior knowledge of the adversary about the deletion patterns of the users becomes more precise. This advantage results in a more accurate adversarial model that lowers the privacy of the users.
\begin{table}[tb]
	\newcolumntype{C}{>{\centering\arraybackslash} m{0.75cm} }
	\newcolumntype{F}{>{\centering\arraybackslash} m{1.cm} }
	\centering
	\small
	\begin{tabular}{|F|C|C|C||C|C|C|}
		\cline{2-7}
		\multicolumn{1}{c|}{}
		& \multicolumn{3}{p{3cm}||}{\textbf{\#FFT (in trillions) for flag-once scenario and diff availability \%}} & \multicolumn{3}{p{3.1cm}|}{\textbf{\#FFT (in trillions) for flag-multi scenario and diff availability \%}} \\ \hline
		{\textbf{DT (days)}} & {\textbf{85\%}}& \textbf{90\%} & \textbf{95\%} & \textbf{85\%} & \textbf{90\%} & \textbf{95\%} \\ 
		\hline\hline
		\multicolumn{1}{|c|}{\textbf{30}} & 1.64 & 1.54 & 1.23 & 13.05 & 8.7 & 4.35 \\ \hline
		\multicolumn{1}{|c|}{\textbf{60}} & 1.45 & 1.24 & 0.83 & 6.39 & 4.26 & 2.13 \\ \hline
		\multicolumn{1}{|c|}{\textbf{90}} & 1.25 & 1.01 &  0.62 & 4.18 & 2.78 & 1.39 \\ \hline
		\multicolumn{1}{|c|}{\textbf{120}} & 1.09 & 0.84 & 0.48 & 3.07 & 2.04 & 1.02 \\ \hline
		\multicolumn{1}{|c|}{\textbf{150}} & 0.95 & 0.71 & 0.40 & 2.40 & 1.60 & 0.80 \\ \hline
		\multicolumn{1}{|c|}{\textbf{180}} & 0.84 & 0.61 & 0.34 & 1.96 & 1.30 & 0.65 \\ 
		\hline
	\end{tabular}
	\vspace{1mm}
	\caption{{Falsely Flagged Tweets (FFT)  with different availabilities, which the adversary needs to investigate under different scenarios. DT denotes decision threshold.}}
	\label{table:false_positives}
\end{table}

	\section{Effect of \system~in Practice}\label{sec:utility}

\noindent Platforms would like to make sure that their users are able to normally interact with the content they want and thus utility of their system is preserved when \system{} is in place.
This guarantee differs from availability since even with 99\% availability, the 1\% non-available content might be the ones users are interested in. 
We identify one key factor that captures the distinction between availability and utility---the interaction with content in many platforms go down with time passing. E.g.,~\cite{vanLiere-2010-retweetTime, sysomos-retweets} shows that tweets receive more than 60\% of their retweets and replies within the first hour of posting and it 
quickly becomes negligible as time passes. 
Thus, in this section we investigate how \system{} preserves the utility and not hinder the normal platform operations. 

\subsection{Quantifying utility of a platform}

\noindent In order to evaluate the effect on utility in a real world scenario, we 
leverage data from Twitter. But first, we need to concretely define the utility of each post as well as the utility of the platform in the context of Twitter.

\noindent \textbf{Utility of a post and of the platform:} 
We take ``retweets'' as a proxy for interactions (temporal utility) around a tweet. 
We quantitatively measure the collective utility of the platform
to be the \textit{fraction of retweets allowed} when \system{} is in place.
Although retweets are only a subset of all interactions 
(other interactions might be replies or user mentions) and may not capture the entirety of interactions,
it is still one of the widely employed proxies of activity around a tweet \cite{mondal-2016-longitudinal-exposure,conover2011political, boyd2010tweet}.

\noindent \textbf{Collecting a utility dataset:} We need to ensure that \system{} preserves utility for all normal users of our system. To create a collective random sample of such users, we first take all the users who appeared in the 1\% random Twitter sample collected in the first week of November 2011. Then we divide the users into five exponential buckets based on their number of followers (i.e. by their popularity) and randomly sampled 500 users from each bucket. We did this subsampling in mid-February 2016. Thus we end up with 2,500 random users. We collected all the tweets posted by these users (respecting Twitter's limit of 3200 most recent tweets per user) and all the retweets of those tweets on end of February 2016. Out of 2,500, 6 users have made their account private between the time of subsampling and the time of all-tweets collection. So we end up collecting data from rest of the 2,494 users. There are a total of 4,858,014 tweets in our dataset. Among them there are 730,055 tweets with at least one retweet and these tweets have 8,836,706 retweets in total. We use this dataset to check the \system's effect on platform utility.

\subsection{How does \system{} affect utility?}

\noindent We simulate \system{} on our utility dataset with the following set-up for \system{}'s parameters.


\noindent \textbf{Setup for measuring utility in presence of \system{}:} We have experimented with setting the platform availability to 85\%, 90\% and 95\%. We again set the mean down time to 1 hour and set mean up times to satisfy the availability requirements. The up and down distributions are geometric and negative binomial respectively. 
Recall that the negative binomial distribution needs a shape parameter along with the mean. Although we are not considering the adversary in the utility experiment, to be consistent with the privacy analysis, we repeat the experiment for the shape parameters from \cref{table:recall_n}.

Specifically, we simulate \system{} for each of the posts in our utility dataset. Note that, an original retweet happening in a down duration (i.e., when the tweet is hidden) is essentially missed and thus platform utility is affected. However, retweets happening in an up duration essentially remain unaffected. We count all the retweets that would have been missed if \system{} was in place and calculate the fraction of retweets missed due to \system{}. Note that, here we do not consider the effect of missed retweets on future retweets, modeling such effect are part of our future work. Finally, the utility of our system will be simply \textit{1 - fraction of retweets missed}.

\begin{table}[t]
	\small
	\centering
	\newcolumntype{C}{>{\centering\arraybackslash} m{1cm} }
	\newcolumntype{F}{>{\centering\arraybackslash} m{2.5cm} }
	\begin{tabular}{|F|C|C|C|}
		\cline{2-4}
		\multicolumn{1}{c|}{} & \multicolumn{3}{c|}{\textbf{Availability}} \\
		\hline
		\textbf{Decision Threshold (days)} & \textbf{85\%}         & \textbf{90\% }        & \textbf{95\%}	\\
		\hline\hline
		\textbf{30}           &	99.25 	& 99.50         & 99.76              \\\hline
		\textbf{60}           &	99.46	& 99.66         & 99.83               \\\hline
		\textbf{90}           &	99.55	& 99.72         & 99.87         		 \\\hline
		\textbf{120}         & 99.61	& 99.76         & 99.89               \\\hline
		\textbf{150}         & 99.63	& 99.79         & 99.90               \\\hline
		\textbf{180}         & 99.68	& 99.82         & 99.91                \\
		\hline
	\end{tabular}
	\vspace{1mm}
	\caption{{Utility for Twitter in presence of \system{} operating with different availabilities and different decision thresholds. In all cases the utility of the system is above 99\%, and as the availability increases the utility increases.}\vspace{-6mm}}
	\label{table:utility}
\end{table}

\noindent \textbf{\system{} has minimal effect on system utility:}  \cref{table:utility} shows the utility of the platform in presence of \system{} with varying decision thresholds (for each of them the optimal shape parameter is used). The table is showing the utility, i.e., the fraction of retweets allowed, for 85, 90 and 95\% availability. For each of the availabilities, we have chosen six different decision thresholds with their corresponding shape parameter from  \cref{table:recall_n}. The key observation is: for all the cases the utility is quite high. Difference between the utilities are at most 0.5\% for different availabilities, and if 99\% utility is sufficient for the platform, the platform can simply choose 85\% availability over 95\% to provide better privacy to the users while maintaining utility.  

In summary, \system{} can indeed hide deletion of users while having minimal effect on platform utility. For a successful \system\ deployment, even 85\% or less availability might provide a good trade off between privacy, availability, adversarial overhead and platform utility.


\section{Enhancements and Discussion}\label{sec:extension}




\paragraph{Real-world restricted adversary} 
In this work, we considered an adversary that can consistently observe each and every post of our \platform{} and has full access to the \system{} up/down distribution parameters.
However, a real-world adversary will have a much more restricted view of the \platform{} (e.g., Twitter normally allows the developers to collect only 1\% random sample of their data) or even of the \system{} deployment (e.g., the adversary has to estimate the exact parameters of up/down distribution). 
Further, in the real world, non-state-level adversaries will be severely limited by computing power and memory. 
Hence a possible extension of \system{} is to restrict the adversary model (i.e., capabilities of the adversary) with practical restrictions on the adversary's resources and  considering the estimation overhead of \system{} parameters. The privacy guarantees provided by \system{} will significantly improve for such restricted, real-world adversaries.

\paragraph{Providing privacy guarantees based on users' needs}
We note that by choosing different up/down time distributions, a \platform{} operator can provide a range of privacy guarantees for \system{}.
For example, if a user needs privacy specifically for 2 or 3 days (e.g., during an uprising) then the system operator can provide short-term privacy by choosing appropriate distributions (where LR value is very low for a short term, then increase rapidly). 
On the other hand, some celebrity might want long term privacy, where the privacy guarantee is not very high, but it is relatively stable over time. In other words, another possible extension will be to match users' need for privacy by simply tweaking the parameters and distributions in \system{}. The privacy guarantees can further be improved in case a user does not mind deleting their content only in down periods. Recall that, we choose geometric distribution as a suitable up distribution primarily since it enables the users to delete their content in both up and down time durations without any effect on the privacy. In case \post{} deletions are restricted only to down durations, we can also explore other choices for up distributions.

\paragraph{Will six month be sufficient?}
\system{} provides plausible deniability guarantees for a deletion even after 3 to 6 months of deletion. 
We argue that delaying an adversary 3 to 6 months to detect deletions might be sufficient in many scenarios. The reason is twofold: (i) Recent work~\cite{DBLP:conf/icwsm/Gomez-RodriguezGS14} modeled users of social platforms as limited memory information processing actors; these actors care less and less about old information. In fact, this model is supported by the phenomena that almost all large social media sites today show the posts in reverse chronologically. (ii) 
Usually, curious people may focus on some specific user's posts related to some offline (i.e., physical work) event (e.g., in the case of the SNL cast member~\cite{SNL}, it was her joining the SNL); however due to the very same reason the user in focus might decide to delete her posts at that time. If \system{} can delay the revelation of this deletion even for a few days, it should be sufficient to dissuade the observers.

\paragraph{Opt-outs and Delayed Execution}
In some cases, users wish to maintain uninterrupted availability of some of their posts infinitely (e.g., pinned tweets on Twitter) or for the first few days. \system{} can easily skip such posts specifically marked by the user. Although these posts do not affect privacy and only improve availability, they can improve adversarial precision: such posts are hardly deleted  and thus, their continuous presence will result in lesser false positives. Nevertheless, given the very high utility provided by \system, we expect the number of such posts to remain limited.

\paragraph{Deception for Intrusion Detection and Surveillance Systems}
\system\ can have interesting applicability beyond the content deletion scenario.
Consider an intrusion detection or surveillance system that continuously monitors 
accesses to a system. Assume an intruder with a side channel that allows him to determine
if the system is not functioning for maintenance, power outage or crash.
The intruder wishes to exploit this side channel to attack the system;
nevertheless, the attack might be time-consuming, and 
the stakes can be very high
such that he does not like to get caught in action. 
\system{}'s approach can be used in this context as a deceptive technology, deterring the 
intruder even when the system goes down.
It will be confusing for the intruder as it cannot determine if the system is in a sleep mode due to \system{} or has crashed.
Interestingly, this approach will also be helpful towards making the surveillance system energy-efficient 
as it will not have to be online and operate constantly.

\section{Conclusion and Future Work}

\noindent In the world with perfect and permanent memory, we are in dire need of mechanisms to restore the ability to forget.
Against an adversary who can persistently observe a user's data, the user's deletions  make her more vulnerable by directly pointing the adversary to sensitive information. In this work, we have defined, formalized, and addressed this problem by designing \system{}. 

In particular, we have formally defined a novel intermittent withdrawal mechanism, 
quantified its privacy, availability, and adversarial overhead guarantees in the form of
a tradeoff. 
We leverage this mechanism to design \system{} which provides users deniability for their deletions 
while having very little impact on the system availability against
an extremely powerful adversary having complete knowledge about the archival \platform{}.
Still, even in the case of such an adversary, 
leveraging real-world data we have demonstrated the efficacy of \system{} in providing a good 
tradeoff between privacy, availability, adversarial overhead and platform utility.
For example, we have shown that while maintaining $95\%$ availability and utility as high as $99.7\%$, 
we can offer deletion privacy for up to $3$ months from the time of deletion 
while still keeping the adversarial precision to $20\%$.

Our work takes first few prominent steps towards solving the multi-faceted problem of forgetting the forgotten, while several interesting challenges remain. 
One future challenge is to consider deletion of correlated posts. 
Another challenge is to handle concrete deployment issues for \system{}, e.g., how to synchronize hiding/unhiding processes between geo-replicated data stores? To conclude, we believe our work calls for further research into these issues in order to provide users a more private right to be forgotten.



	\paragraph{Acknowledgment}
\noindent The authors gratefully thank the anonymous reviewers for their helpful comments. 
This work was partially supported by an Intel-CERIAS research fellowship and the Alexander von Humboldt foundation.

	\bibliographystyle{IEEEtranS}
	\bibliography{content_deletion}

	\appendix

\subsection{Difference between Deletion Privacy \& Differential Privacy}\label{appendix:diff}

Our notion of deletion privacy has parallels with differential privacy~\cite{Dwork-2006-dp} in that we consider the ratio of likelihood of observed states, but there is also a subtle difference. The privacy parameter defined in Definition~\ref{def.privacy} depends on the specific observed states $\nobs$. This is in contrast with differential privacy where the relevant ratio $e^\epsilon$ (for the parameter $\epsilon$) is defined as a worst-case bound \emph{for all} possible observations. 
The reason for choosing this definition instead of differential privacy is that 
it is not possible to find a meaningful bound on the ratio in Equation~\eqref{eq:main_eq} 
valid for all observations: as time-since-deletion increases, the adversary becomes more certain about deletion. 
In short we can interpret our deletion privacy definition as a way to capture the certainty of an adversary for detecting \post{} deletion 
{\em with} his observed states over time.

\subsection{Effect of Negative Binomial Shape Parameter}\label{section:shape_param_disscussion}

\noindent \textbf{What is a suitable shape parameter for negative binomial distribution?} 
We present an analytical approach to set an optimal $n$, given \platform{} operator's estimate of decision threshold $\decisionThresEst$. Since we set the up time distribution as geometric distribution with a constant inverse hazard rate (which we will denote by ``$c$''), Equation~\eqref{eq:main_res} becomes

$
LR =  
\dfrac
{ c + 1}
{\ccdfDo{\downDur - 1 }}.
$

Ideally, the \platform{} operator should set $n$ such that, when the adversary's decision threshold is $\decisionThresEst$ (i.e., the adversary flags a \post{} as deleted after not observing the post for time $\decisionThresEst$ or more), the post has the lowest LR value. In other words, LR value should be lowest when the last down duration is $\decisionThresEst$. Thus, by deciding negative binomial distribution with mean ${\mu_d}$ and shape parameter $n$, we would want  $\ccdfDo{\downDur - 1}$ to reach a maximum at $\downDur = \decisionThresEst$. Thus, we take the derivative of $\ccdfDo{\downDur - 1}$ with respect to shape parameter $n$ and equate it to 0 at $\downDur = \decisionThresEst$, i.e.,

\begin{equation}
\label{eq:driv}
\frac{\partial}{\partial n}\ccdfDo{\decisionThresEst - 1} = \frac{\partial}{\partial n} I_{\left(1-n\frac{1-\mu_d}{\mu_d}\right)}(\decisionThresEst, n) = 0   
\end{equation}
where $I_x(a,b)$ is the incomplete beta integral. Now setting ${\mu_d}=1\textrm{ hour}$, we solve for $n$ to determine the best shape parameter for a given value of $\decisionThresEst$.

\begin{table}[tb]
	\small
	\newcolumntype{F}{>{\centering\arraybackslash} m{3cm} }
	\newcolumntype{C}{>{\centering\arraybackslash} m{.45cm} }
	\begin{center}
		\begin{tabular}{ |F||CCCCCC|}
			\hline
			\textbf{Estimate of decision threshold is $\decisionThresEst$ days}& 30 & 60 & 90 & 120 & 150 &180\\ 
			\hline
			\textbf{Shape parameter $n$ for lowest LR $\times 10^{-4}$} &6 & 3& 2& 1.5& 1.2& 1\\
			\hline
		\end{tabular}
		\vspace{2mm}
		\caption{{The best shape parameter $n$ i.e. the lowest LR value when the estimated decision threshold for the adversary is $\decisionThresEst$ days. The mean of our negative binomial distribution is one hour.}}\label{table:recall_n}
	\end{center}
	
\end{table}

\cref{table:recall_n} shows the best shape parameters for different values of $\decisionThresEst$. An archive operator can choose any of these values according to her choice of $\decisionThresEst$ or even calculate suitable values of $n$ for her estimated $\decisionThresEst$ using our analytical technique. 

\begin{figure}[h]
	\centering
	\includegraphics[width=0.85\columnwidth]{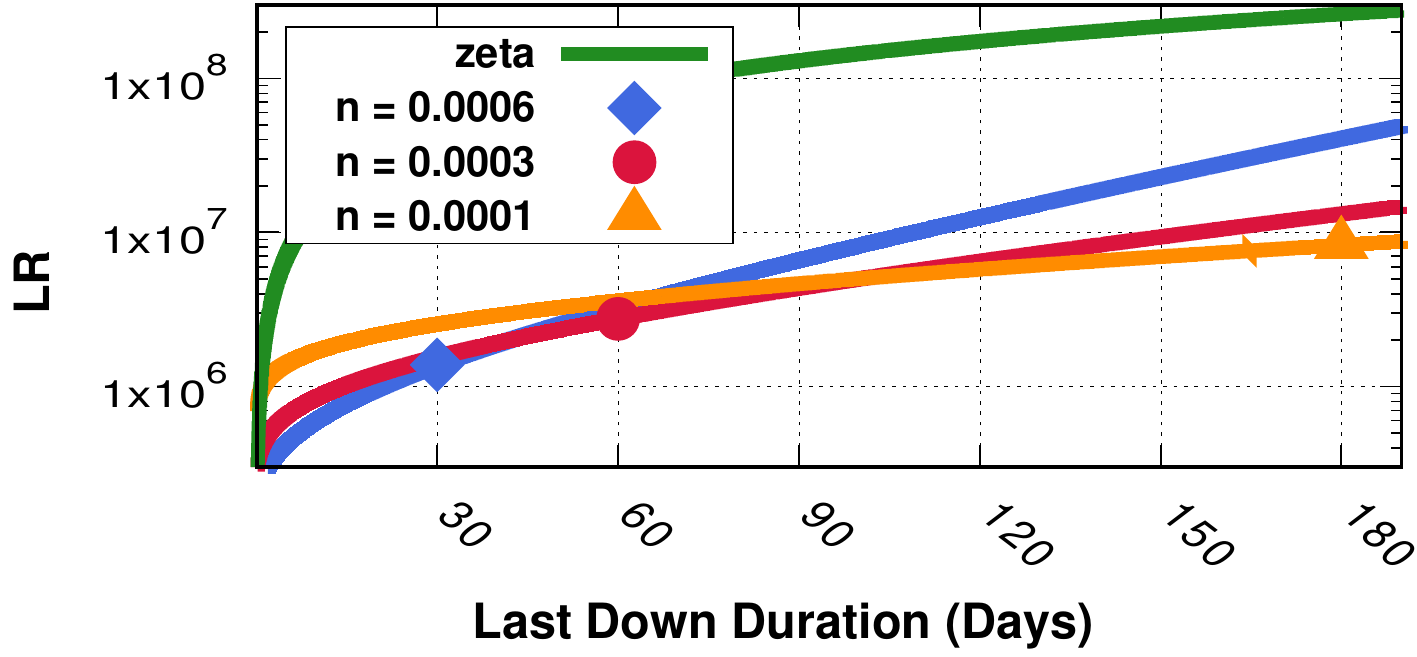}\vspace{-2ex}
	\caption{\textbf{Variation of LR for zeta distribution and three choices of shape parameters for the negative binomial distribution. The x-axis is showing the last down duration in months and y-axis is showing the LR value in log scaled. The lowest LR value for each of the decision thresholds in table \ref{table:recall_n} is for the corresponding shape parameter. Choosing negative binomial distribution with any of the parameters for the given down durations results lower LR values then choosing zeta as the down distribution.}}\vspace{-1mm}
	\label{fig:lr_different_shape_param}
\end{figure}

\noindent \textbf{Effect of Negative binomial Shape Parameter on LR value:} Furthermore, in \cref{fig:lr_different_shape_param}, we demonstrate how parameter $n$ impacts the LR value by plotting the LR for some of the shape parameters in table \ref{table:recall_n} (corresponding to $\decisionThresEst$ 1, 2 and 6 months). 
We have set the mean up and down time to nine and one hours respectively.
As evident there is no shape parameter that performs best for all the times, 
however, we can observe that for each of the decision thresholds $\decisionThresEst$ in  \cref{table:recall_n} 
the corresponding shape parameter has the lowest LR value. 
We also observe that for all the chosen parameters, LR value for negative binomial is lower than the case of picking zeta as down distribution.

\section{An Overview of \system\ Implementation}\label{implementation}

\noindent In section~\ref{sec:Design} we presented the basic steps of \system{}. However in our paper we considered that \system{} should be applied to each single post. So a very pratical question is: How to efficiently implement \system{} in a platform? Here we provide a brief implementation sketch.

\vspace{1mm}

\paragraph{Basic setup for a platform}We assume a generic archival platform where each \post{} is stored as an Active Store Object (ASO)~\cite{geambasu-2010-comet}. ASOs are simply key-value pairs with some (optional) code to run on values. Traditionally this ASO code is written in terms of handlers (e.g., code to handle deletion). Each \post{} ASO will have an unique \post{} id as key, the user generated \post{} content as value, identification of the owner (as authentication token) and some metadata (e.g., the real state flag for a \post{}). We further assume that there is an internal trusted time server, which is used throughout the \platform{} for synchronizing operations. The \platform{} internally does not use any other timestamps. Any mention of timestamps in this section refers to this internal timestamp. Extending this set-up to traditional databases is simple and left to future work.


We use an architecture similar to Comet~\cite{geambasu-2010-comet}, where the platform operator as well as platform users (including adversary) have some specific Application Programmer Interfaces (API) to access/create/delete the \posts{}. However note that in our adversary model, the adversary can just query the \posts{} and can not change them in any way. Thus, unlike Comet in \system{} \post{} ASO objects are immutable from the point of view of an adversary.


\vspace{1mm}
\paragraph{Straw man implementation} A straightforward implementation of \system{} is to add an ``observable state'' flag (binary) with meta data of each \post{} ASO. Whenever a \post{} is created, the \platform{} operator assigns a process (or a thread) to the \post{}. That process will apply \system{} algorithm to toggle the observable state. In case of a view request, another user initiated process will seek the required ASO or ASOs, check the ``observable state'' flag and return a \post{} if the \post{} is observable (i.e., observable state flag is \textsf{TRUE}). However, this design if definitely not scalable for a platform with billions of \posts{}. Thus we need an improved implementation.

\vspace{1mm}
\paragraph{Key insight}Our key insight is simple---the \platform{} can precompute the timestamps for future up and down durations and then lazily update those duration timestamps. At any current time, for a view request, the platform operator can use the current timestamp to determine if the post should be in up or down duration (using the precomputed up/down duration timestamps) and return a post in case the post is in up duration or return null otherwise. The only exception is if the data owner requested to view her own post, the post should be returned, irrespective of up/down duration.

\begin{figure}[t!]
	\centering
	\includegraphics[width=0.9\columnwidth]{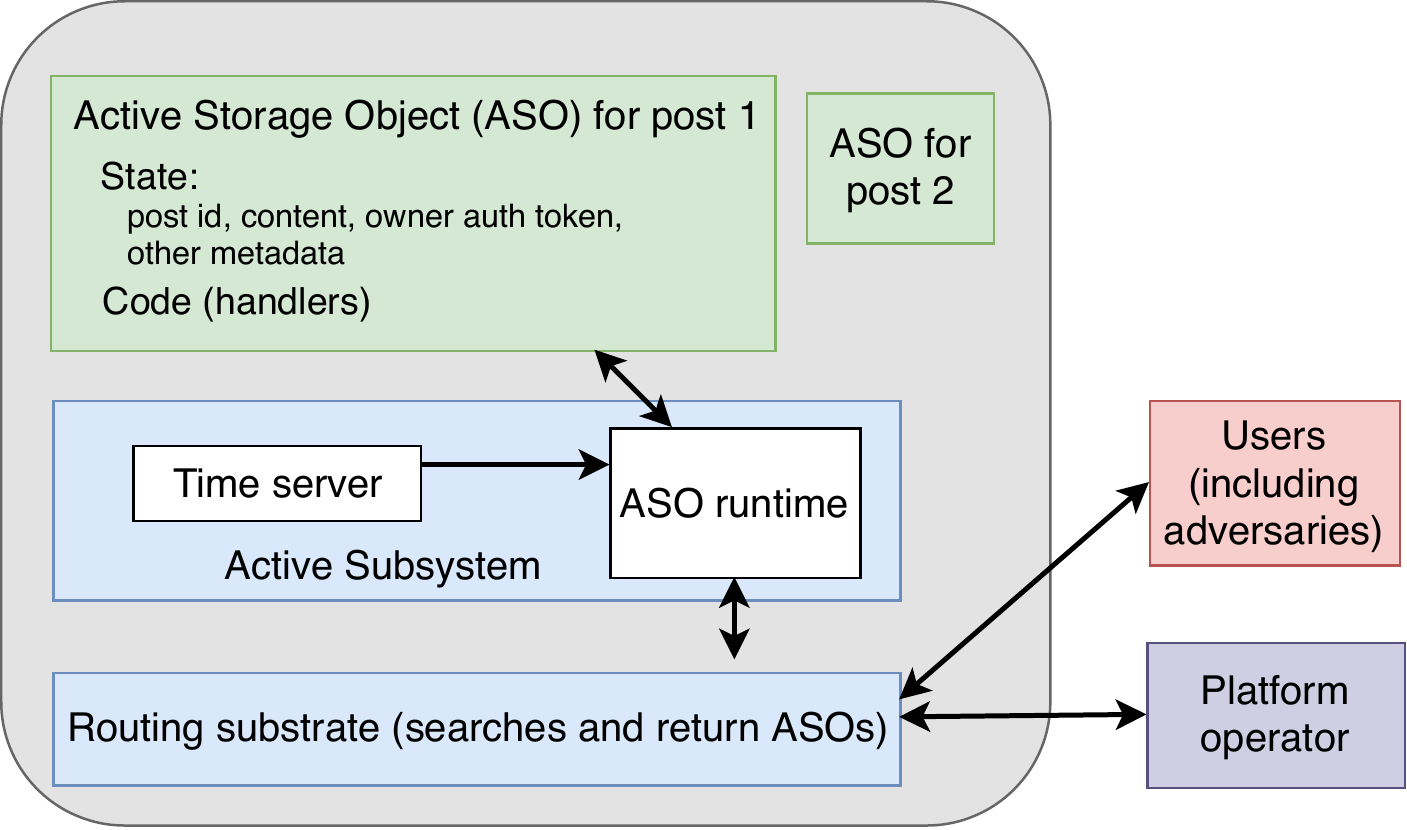}
	\caption{\textbf{A basic implementation schematic for \system{}. Each post is an ASO, and using APIs and code handlers these ASOs can be accessed. An operator can add more metadate to the ASO content according to requirement of the \platform{}.}}\vspace{-4mm}
	\label{fig:letheia_impl}
\end{figure}

\vspace{1mm}
\paragraph{An improved \system{} implementation} We note that instead of keeping track of the observable state, a process can simply compute the observable state of an ASO using the current timestamp and the precomputed up/down duration timestamps. Thus, when the \platform{} operator adds each \post{} ASO, they should also add (in bulk) the timestamps corresponding to up and down durations for a large time period in future (e.g., for next one year). Specifically we present the basic schematic of our proposed implementation in Figure~\ref{fig:letheia_impl}. Each \post{} ASO contains a state which includes the \post{} content, an authentication token to identify the owner of the \post{} (who can delete the \post{}) and timestamps for future up and down durations. Both users (including adversaries) and \platform{} uses a routing substrate to find ASOs in the distributed storage (e.g., via a hash table of keys). The active subsystem contains a trusted time server and the ASO runtime, which converts \platform{} and user API calls to ASO handlers and executes the ASO handler code.

\begin{table}[t]
	\centering
	\footnotesize
	\begin{tabular}{|p{1.1cm}|p{2.1cm}|p{3.3cm}|p{1.1cm}|}
		\hline
		\multicolumn{4}{|p{7.6cm}|}{\textbf{APIs for the user (including adversary)}} \\ \hline\hline
		\textbf{Name} &\textbf{Parameter} & \textbf{Description} & \textbf{Associated ASO handlers} \\ 
		\hline
		$put$ & post\_content, authentication\_token & Creates a post ASO for the data owner, returns a post\_id  & $onPut$ \\ \hline
		$get$ & post\_id, authentication\_token & Returns a post ASO or null depending on  (i) ownership and (ii) if the post is in up/down duration.  & $onGet$ \\ \hline
		$delete$ & post\_id, authentication\_token & deletes associated post and returns null.  & $onDelete$ \\ \hline
		\hline
		\multicolumn{4}{|p{7.6cm}|}{\textbf{Internal APIs for the \platform{}}} \\ \hline\hline
		\textbf{Name} &\textbf{Parameter} & \textbf{Description} & \textbf{Associated ASO handlers} \\ 
		\hline
		$updateTS$ & list of post\_ids to update & Updates the future up/down times in  ASOs with post\_ids.  & $onUpdate$ \\ \hline
		\hline
		\multicolumn{4}{|p{7.6cm}|}{\textbf{Handlers in ASOs}} \\ \hline\hline
		\textbf{Name} &\textbf{Parameter} & \textbf{Description} & \textbf{Associated ASO handlers} \\ 
		\hline
		$onPut$ & post\_content, authentication\_token, current\_timestamp & Creates an ASO object and assigns up/down timestamps covering next 1 year.  & - \\ \hline
		$onGet$ & post\_id, authentication\_token, current\_timestamp & Check current timestamp and if in up duration return post content, else return null.  & - \\ \hline
		$onDelete$ & post\_id, authentication\_token, current\_timestamp & Assign one down timestamp---infinity; remove post content. & - \\ \hline
		$onUpdate$ & post\_id, authentication\_token, current\_timestamp & If current set of up/down times cover less than 1 year, create more up/down times. & - \\ \hline
				
	\end{tabular}
	\vspace{1mm}
	\caption{Summary of API and ASO handler code descriptions (and the mapping between them) for \system{} sketch implementation. Note that data owner always gets back her non-deleted posts irrespective of up/down duration.}\vspace{-8mm}
	\label{table:letheia_sketch_api_handler}

\end{table}

Table~\ref{table:letheia_sketch_api_handler} contains the summary of API and ASO handler code descriptions. We use authentication (or auth) tokens to identify a user (to determine data owner or not). Any user can create a \post{} using her auth token with $put$ or delete her posts using $delete$. Handler code for call $get$ first checks the auth token and if the request is from data owner the \platform{} always returns the \post{} (if it is not deleted). If the $get$ request is not from a data owner, then (using the precomputed up/down durations) the handler code checks if the current timestamp is within the up of down time duration. If the current timestamp falls in an up duration for the \post{} then the \platform{} returns the \post{}'s content to the requesting user, otherwise the \platform{} returns null. In addition to $get$, $put$ and $delete$, the \platform{} operator internally runs multiple processes with $updateTS$ function to keep adding future up and down time durations for ASO objects. The ``post\_ids'' to update (given to $updateTS$) should be divided in these processes based on a hash table of ASO keys. The mapping between API and ASO handler codes is in $3^{rd}$ column of Table ~\ref{table:letheia_sketch_api_handler}.

\vspace{1mm}
\paragraph{Possible optimizations of this implementation sketch}We emphasize that this is just a sketch \system{} implementation with scopes for further optimization.
E.g., $updateTS$ can additionally delete up/down timestamps lesser than current timestamp to optimize storage or there can be batch garbage collection after multiple calls to $delete$. Further the input to $updateTS$ can be chosen more intelligently
e.g., by keeping a min-heap to determine the ASO objects which are in immediate need to update up/down timestamp. We leave exploration of these concrete system challenges to future work. 


\end{document}